\documentclass[fleqn,usenatbib]{mnras}

\usepackage{newtxtext,newtxmath}
\usepackage{float}

\usepackage[T1]{fontenc}
\usepackage{graphicx}
\usepackage{epstopdf}
\usepackage{natbib}
\usepackage{hyperref}
\hypersetup{colorlinks,citecolor=blue}
\usepackage{etoolbox}
\usepackage{rotating}
\usepackage{pst-node}
\usepackage{soul} 

\usepackage{pstricks}
\usepackage{pst-all}
\usepackage{ifpdf}
\usepackage{amsmath}	
\usepackage{hyphenat}
\emergencystretch=\maxdimen
\hyphenchar\font=-1
\title[The interacting system AM\,1204-292] 
{The physical properties and the evolution of the interacting system AM\,1204-292} 
\author[D. A. Rosa et al.]{
D. A. Rosa,$^{1}$\thanks{E-mail: deiserosa@univap.br}
I. Rodrigues,$^{1}$
A. C. Krabbe, $^{1}$
A. C. Milone,$^{2}$
and S. Carvalho $^{1}$
\\
$^{1}$Instituto de Pesquisa \& Desenvolvimento (IP\&D), Universidade do Vale do Para\'iba, 
             Av. Shishima Hifumi 2911, 12244-000,  S\~ao Jos\'e dos Campos-SP, Brazil.\\
$^{2}$Instituto Nacional de Pesquisas Espaciais (INPE), Divis\~ao de Astrof\'isica, 
             Av. dos Astronautas 1758, 12227-010, S\~ao Jos\'e dos Campos-SP, Brazil.
}

\date{Accepted 2020 December 07. Received 2020 December 06; in original form 2019 December 09}

\pubyear{2020}

\begin{document}
\label{firstpage}
\pagerange{\pageref{firstpage}--\pageref{lastpage}}
\maketitle

\begin{abstract}

We investigate the interaction effects in the stellar and gas kinematics, stellar population and ionized gas properties in the  interacting galaxy pair AM\,1209-292, composed by NGC\,4105 and NGC\,4106. The data consist of long-slit spectra  in the range of $3000-7050$\,{\AA}. 
The massive E3 galaxy NGC\,4105 presents a flat  stellar velocity profile, while the ionized gas is in strong rotation, suggesting external origin. Its companion, NGC\,4106, shows asymmetries in the radial velocity field, likely due to the interaction.  The dynamics of the interacting pair was modelled using P-Gadget3 TreePM/SPH code, from which we show that the system has just passed the first perigalacticum, which triggered an outbreak of star formation, currently at full maximum.
We characterized the stellar population properties using the stellar population synthesis code {\scriptsize\,STARLIGHT}  and,  on average, both galaxies are predominantly composed of old stellar populations.
NGC\,4105 has a slightly negative age gradient, comparable to that of the most massive elliptical galaxies, but a steeper metallicity gradient. The SB0 galaxy NGC\,4106 presents smaller radial variations in both age and metallicity in comparison with intermediate mass early-type galaxies.
These gradients have not been disturbed by the interaction since the star formation happened very recently and was not extensive in mass. 
Electron density estimates for the pair are systematically higher than those obtained in isolated galaxies. The central O/H abundances were obtained from  photoionization models in combination with  emission line ratios, which  
resulted in  $12+{\rm log(O/H)}=9.03\pm0.02$ and  $12+{\rm log(O/H)}=8.69\pm0.05$ for NGC\,4105 and NGC\,4106, respectively.

\end{abstract}




\section{Introduction}
\label{sec:introduction}
Interacting pairs of galaxies are excellent laboratories
to study different aspects of the galaxy evolution, such as the  morphological transformation, the  induced star formation, and the fueling of a central supermassive black hole (lighting an active nucleus) \citep[e.g.,][]{1995PASP..107.1129F,1995A&A...297..331D,1996A&A...308..387D,2000A&A...353..917L,2008A&A...484..655F,2009MNRAS.399.2172R}.
Besides, the study of binary galaxies at different interaction stages can provide important constraints about the hierarchical scenario of galaxy formation and evolution.
The relatively small velocities of a close encounter in a gravitationally coupled pair of galaxies of comparable masses/sizes make the mutual interaction strong enough inside each galaxy. Numerous studies have reported that a close encounter between the two galaxies modify the mutual gravitational fields \citep{1972ApJ...178..623T}.
A  close encounter can create new substructures in the gas, dust, and star spatial distributions,
such as warps or bars, tidal tails, plumes, and bridges, pulled out by tidal forces during the gravitational interaction
\citep[e.g.,][]{1994ApJ...427..684M, 1999AJ....117.2695R, 2002MNRAS.333..481B,2010PhDT.........1R, 2010A&A...510A..31P, 2011A&A...533A..22D}.

Mixed pairs of galaxies have components spanning different morphologies,
such as an elliptical (early-type galaxy) and a spiral (late-type galaxy). This kind of galaxy pair becomes 
particularly more interesting because they represent good sites for investigating in detail those physical 
processes induced by close interactions \citep[e.g.,][]{1995PASP..107.1129F, 1995A&A...297..331D, 1996A&A...308..387D}. 
The Catalogue of Southern Peculiar Galaxies and Associations by Arp and Madore \citep{arp1987catalogue} became a pioneer 
compilation of galaxy pairs, in which a lot of mixed pairs can be found.

The dynamics of galaxy interactions can be studied through numerical simulations, including the stellar and gaseous 
components besides the dark matter as primary ingredients. 
For instance, \citet{2015MNRAS.448.1107M} embarked on a
systematic study of 75 simulations of galaxies in major mergers.
They report that star formation is more elevated in the center than the ones on the
outskirts of interacting galaxies, and concluded that these trends are most prominent in the
smaller companion galaxies with strongly aligned disc spin orientations.
Another example is the numerical simulations by \citet{2012ApJ...746..108T}
that showed that the nuclear metallicity evolution is a perfect competition between the metal-poor gas inflows 
and enrichment due to the star formation (or feedback from star formation and active galactic nuclei (AGN) activity).

Mutual interaction between galaxies may enhance star formation \citep[e.g.,][]{1978ApJ...219...46L, 2007AJ....134..527W, 2008MNRAS.389.1593K, 2013MNRAS.433L..59P, 2015MNRAS.454.1742K}.
However, the environment might not play a determinant role
in the formation and evolution of (very) massive early-type galaxies
\citep[e.g.,][]{xu2010, 2017MNRAS.471.2687B,2018JApA...39...31Y}.

The star formation enhancements are accompanied by other events such as perturbations
in the radial velocity field and dilution of the metallicity gradient.
This is consistent with results obtained in chemical abundances and mass-metallicity relation \citep{2013A&A...554A..58S},
which show that interaction-induced flows of low metallicity gas from the outer parts of the disk of a galaxy
can decrease the metallicity in the inner regions and modify the radial abundance gradients
\citep[e.g.,][]{2010ApJ...721L..48K, 2014MNRAS.444.2005R, 2014A&A...563A..49S, 2018MNRAS.tmp.1838M}.
We can not exclude the possibility that the inflow of low-metallicity gas would be considered as the principal source that affects the nuclear metallicity of interacting galaxies.
Although there is evidence for a dilution in the central gas metallicity produced by interactions,
a few interactions may also be able to enhance it
\citep[e.g.][]{2015A&A...579A..45B},
observationally seen by redder optical colours
\citep[e.g.,][]{2010MNRAS.401.1552D, 2011MNRAS.412..591P}, for instance.

Strong interactions in close pairs may produce an in-homogeneity in the interstellar medium
shown as a wide variation of the electron density \citep[e.g.,][]{2014MNRAS.437.1155K, 2015MNRAS.453.2349R}. 
Formation of bars induced by the perturbation from the companion is also observed \citep[e.g.,][]{2012ApJ...761L...6M, 2018A&A...618A.149A}.

Active galactic nuclei can also be produced, fueled by the infall of gas from the galaxy outskirts towards the central region, or even from a late-type galaxy to its companion
\citep[e.g.,][]{2011A&A...533A.104E, 2011MNRAS.418.2043E, 2014MNRAS.441.1297S, 2018ApJ...853...63D}.
The gravitational interactions, as well as the tidal effects, can provide fuel to a supermassive black hole in the center of the galaxy 
(nucleus), leading to a connection between star formation and black hole activity \citep{2011ApJ...742...46T, 2011ApJ...743....2S}.
The activity of the nucleus provides energy to the interstellar medium and can even smooth the star formation 
\citep{2006MNRAS.365...11C, 2008A&A...492...31D, 2019MNRAS.485.3446H}.
However, \citet{2018A&A...614A..32W} who investigated the effects of
galaxy merger throughout the interaction sequence in NGC\,7252 (nearest major-merger galaxy remnants), revealing the extent of ongoing star 
formation and recent star formation history.
These authors also found a large ionized gas cloud previously discovered $\sim5$\,kpc south of the nucleus, which may be associated with a 
low-ionization nuclear emission-line region (LINER).
Therefore, it is of great significance to study the properties of ionized gas in the circumnuclear region and AGN connected with interacting galaxies in order to evidence the relationship between the active galactic nucleus and gravitational interaction.

This paper presents a detailed study for the interacting system AM\,1204-292,  where the physical properties of the ionized gas, 
the characteristics of the stellar population, and the nuclear activity are linked with the process of the interaction between the galaxies 
and the destiny of this system is uncovered. 

\begin{figure*}
\centering
	\includegraphics[width=\textwidth]{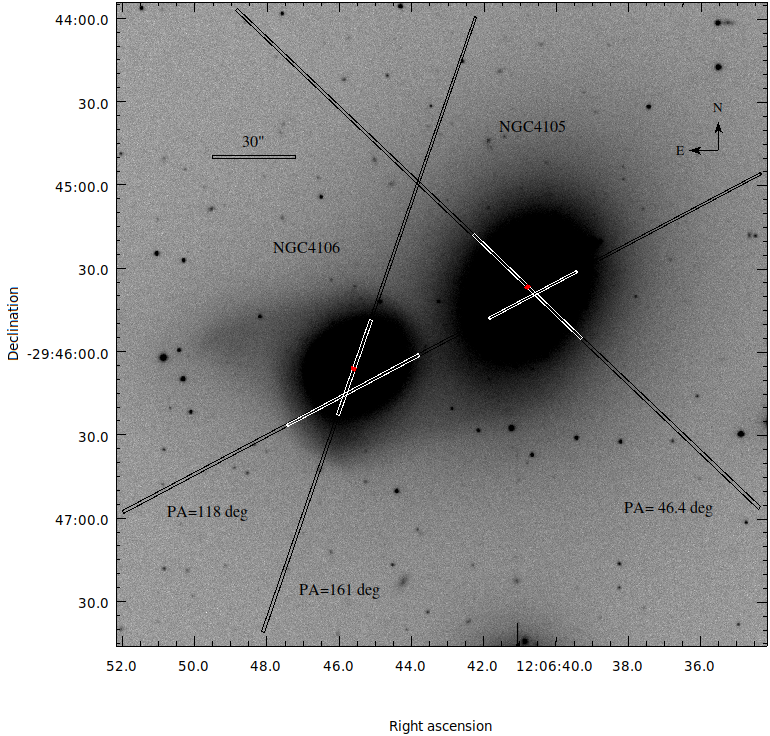}
    \caption{Slit positions for AM\,1204-292 system superimposed on the optical image of the pair in $I$-band from  the Carnegie-Irvine Galaxy Survey. The center of each galaxy is marked in red. The white section of each slit shows the section of the slits from where useful data was extracted \citep{2011ApJS..197...21H,2011ApJS..197...22L}.}
    \label{fig:par}
\end{figure*}

\begin{table*}
	\centering
	\caption{Main characteristics of the AM\,1204-292 galaxy pair members: identification, morphology, total B-band luminosity, colour index (B-V), effective radius R$_e$, apparent major axis direction MA, and inclination angle $i$.}
	\label{tab:sample}
	\begin{tabular}{lcccccccr} 
		\hline
ID &Morphology\,$^{[1]}$ & {\bf log $L_{B}\,~^{[6]}$} & (B-V)$\,^{[2]}$ &   R$_{\rm e}$\,$^{[3]}$ &MA\,~$^{[2,4,5]}$&\textit{i}\,~$^{[2,5]}$ &   others designations \\
AM\,1204-292 &                     & ($L_{\odot}$) & (mag) & (kpc) & ($^{\circ}$)      &($^{\circ}$) &   &  \\  

        \hline
NGC\,4105 & E3       & 10.86 & 0.99 & 10.26 & 136.4 & 54.7 & ESO440-054 \\
NGC\,4106 & SB(s)0+  & 10.75& 1.01 & 3.47  & 91    & 50   & ESO440-056 \\
        \hline
    \end{tabular}
\begin{minipage}[l]{18cm}
{\it References:} 
[1] \citet{1991S&T....82Q.621D}; 
[2] Hyperleda\,\citep{2014A&A...570A..13M}; 
[3] \citet{2015ApJS..219....4S};
[4] \citet{2015A&A...581A..10C}; 
[5] \citet{2015A&A...582A..86H}; 
[6] \citet{2001A&A...374..454T}.\\
\end{minipage}	
\end{table*}

\begin{table}
	\centering
	\caption{Galaxy ID, slit position, exposure time, and wavelength range of the spectra on galaxy pair.}
	\label{tab:observed}
	\begin{tabular}{lccr} 
		\hline
		galaxy\,ID                 & PA\,($^{\circ}$) & Exposure time\,(s)  & $\Delta\lambda$\,({\AA})\\
		\hline
		NGC\,4105                 & 46.4           & 3$\times$1200         & 3000-7050\\
		NGC\,4106                 & 161            & 3$\times$1200         & 3000-7050\\
		NGC\,4105/06              & 118            & 2$\times$1200         & 3000-7050\\
		\hline
	\end{tabular}
\end{table}

AM\,1204-292 is pair of galaxies with mixed morphology components, containing the very luminous elliptical galaxy NGC\,4105 \citep[according to the Third Reference Catalogue  of  Bright Galaxies][hereafter RC3]{1991S&T....82Q.621D}, but 
classified  as  S0  in  the  Revised  Shapley Ames Catalog of Bright Galaxies 
(\citet{1981rsac.book.....S}, hereafter RSA) and NGC\,4106, a (diffuse) lenticular barred galaxy with extended tidal arms produced by the interaction \citep{2000A&AS..145...71K}. It was classified in RC3 as SB0(s)+, and as SB0/a in RSA (tides).

The galaxies are separated by a projected distance of 9.6\,Kpc 
and a difference in the radial velocities of 274\,km s$^{-1}$.
There is visual evidence for one tidal arm in NGC\,4106 that expands towards the primary galaxy in the south direction, 
and its counterpart tidal arm/tail heading to the opposite direction. On the other side,  
NGC\,4105  does not show clear signs of perturbation. 
The spectra of both galaxies are dominated by prominent absorption features \citep{2000AJ....120..189D} 
and some  emission lines, such as [O{\scriptsize\,II}]$\lambda$3727\,{\AA}
\citep{1998A&AS..130..267L} and [N{\scriptsize\,II}]$\lambda$6584\,{\AA} \citep{2000ApJS..127...39C}.  Moreover, \cite{1998A&AS..130..267L}
found evidence from the line-strength indices that these galaxies
experienced recent star formation episodes.
Both galaxies are also fairly red with (B$-$V) color index very similar to $\sim+0.9$ \citep{1996A&AS..116..515R}.
The X-ray luminosity of 
NGC\,4105 is  L$_{x}\sim3.98\times10^{39}$\,erg\,s$^{-1}$\,(0.5\,–\,2.0 keV), and NGC\,4106 is L$_{x}\sim1.26\times10^{39}$\,erg\,s$^{-1}$ \citep{2007AJ....133..220G}. Besides, 
\citet{2009A&A...502..473G} found dusty features in the central region and modeled the brightness distribution, concluding that it presents a disky outer structure.

The paper is organized as follows.
In Section~\ref{sec:obs}, we describe the observations and data reduction.
The stellar and gas kinematics in each individual galaxy are presented in Section~\ref{sec:cine}.
The N-Body\,+\,SPH simulations built to reproduce the interaction between the two galaxies and to uncover
the destiny of this system are explored in Section~\ref{sec:simulations}.
We present in Section~\ref{sec:sps} the procedure employed to perform the stellar population synthesis
and the respective results to estimate the age distribution of the stellar populations.
In Section~\ref{sec:gas_proprie}, we analyze the results about the ionized gas properties, 
centering on electron density and oxygen abundance.
Finally, we summarize our conclusions in Section~\ref{sec:conclusions}.

\section{Observations and data reduction}
\label{sec:obs}

In Table~\ref{tab:sample} some general characteristics of the galaxy pair are listed,  such as:
identification, morphology, absolute $B$-band magnitude, 
effective radius R$_{\rm e}$ (isophotal radius that contains half of the total flux of the component),
major axis  direction (MA), and the inclination angle (\textit{i}) of each galaxy. 
The projected linear separation between the galaxy nuclei on the sky plane  is about 9.6\,kpc 
(adopting $H_{0}=73\,\rm km\:s^{-1} Mpc^{-1}$ \citep{2006PASP..118.1711W},
and the average distance of the galaxy pair of 28.6\,Mpc).

The current study is based on long slit spectroscopic observations carried out on 2017 March 04$^{th}$ 
with the Goodman High Throughput Spectrograph (GTHS)
attached to the 4.10m SOAR telescope (Cerro Pachon, Chile).
Spectra with the single long slit mask were acquired in the range $\lambda\lambda$3000-7050\,{\AA}
with a grating of 400 lines\,mm$^{-1}$ (M1 mode) and a slit of 1.03\,arcsec width,
providing  an average spectral resolution of 5.5\,{\AA} in the region $\lambda\lambda$3800-6800\,{\AA},
a spectral sampling of 1\,{\AA}\,px$^{-1}$, and a spatial scale of 0.15\,\arcsec\,px$^{-1}$. 
The estimate of the spectral resolution
was based on the broadening of He-Ar lines in the wavelength calibration spectra and
cross correlations between the stellar template spectra
(representing $\sigma_{v,inst}$\,=\,132\,km\,s$^{-1}$). The atmospheric seeing was about 1\,\arcsec, and we also noted some 
cirrus clouds along the observations.

The long slit spectra of the galaxy pair AM\,1204-292  were taken at three different position angles (PA) in order to observe the central regions and the principal-axes.
PA$=118^{\circ}$  is crossing the disc of NGC\,4105, nearby the semi-major axis, and NGC\,4106, but not in their  centers.
The minor photometric axis of the elliptical galaxy NGC\,4105 was observed with PA\,=\,46.4$^{\circ}$.
The slit  at PA$=161^{\circ}$  crossed  the center  of NGC\,4106 and the southern tidal arm that extends toward NGC\,4105.

Multiple spectra were taken at the same slit position to increase the signal to noise ratio, and the  exposure times were limited to 1200 seconds to minimize the effects of spurious cosmic rays.
The slit positions are shown in Fig.~\ref{fig:par}, superimposed on the optical image of the pair.
The journal of observations  is listed in Table~\ref{tab:observed}.

The spectroscopic data have been reduced with the {\scriptsize\,IRAF}\footnote{Image Reduction and Analysis Facility, published by National Optical Astronomical Observatory ({\scriptsize\,NOAO}), operated by AURA, Inc, under agreement with NFS.} package. The standard spectroscopic reduction procedure has been applied: image trimming, bias subtraction, flat field correction,
sky background subtraction, cosmic rays removal, one-dimensional spectrum extraction, and wavelength and flux calibrations.
Each one-dimensional spectrum comprises the flux contained in an aperture of 1.03\,$\times$\,0.9\,\arcsec, which at a distance to the par corresponds to nearly 142\,$\times$\,127.5\,pc. 
The galactocentric radial coordinates to the positions of each aperture extracted from the slits were corrected taking into account the inclination angles of each galaxy (given in Table~\ref{tab:sample}), in order to make radial plots of quantities of interest. 

\begin{figure}
    \centering  
	\includegraphics[angle=270,width=0.85\columnwidth]{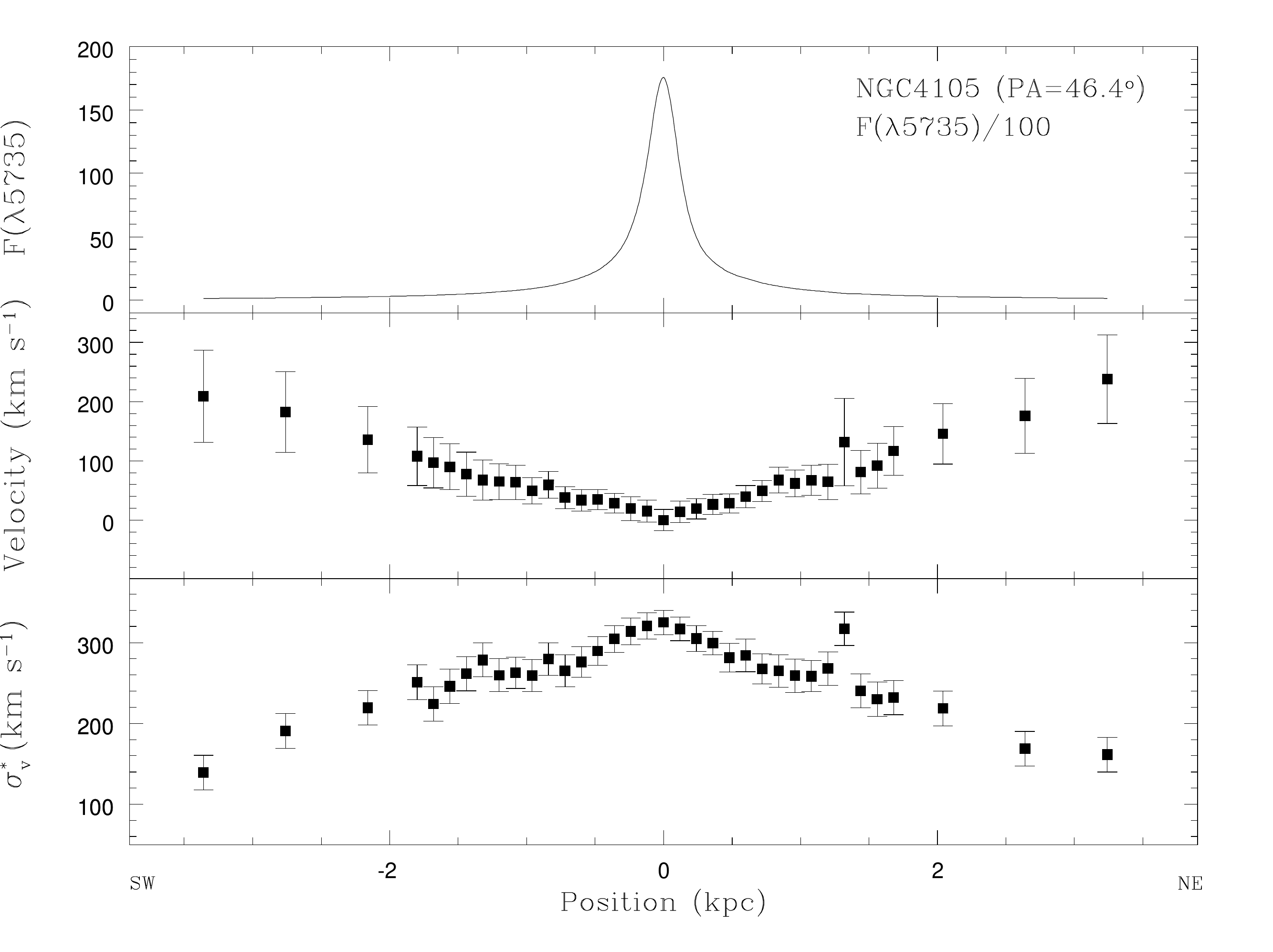}
	\includegraphics[angle=270,width=0.85\columnwidth]{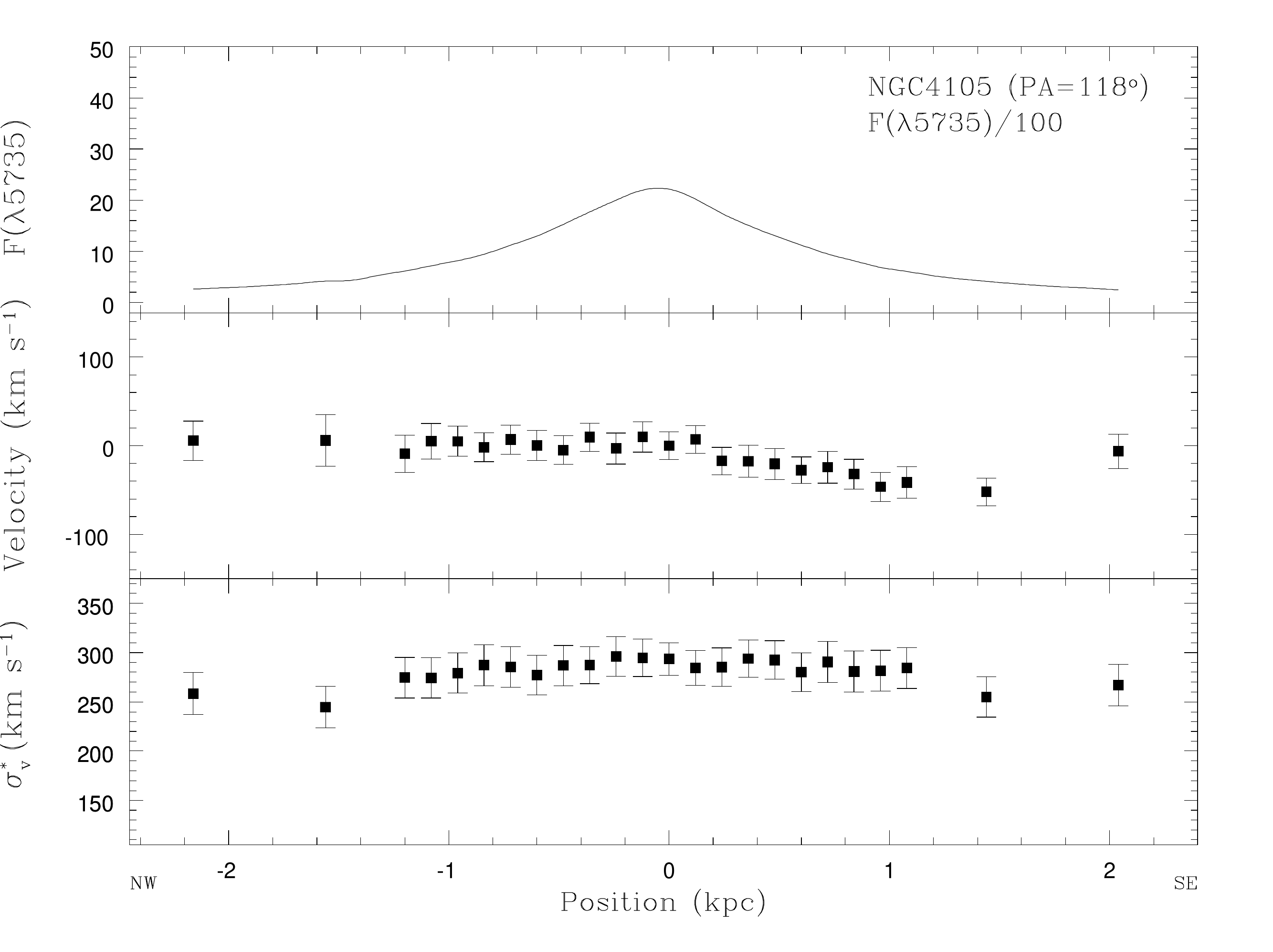}	
	\includegraphics[angle=270,width=0.85\columnwidth]{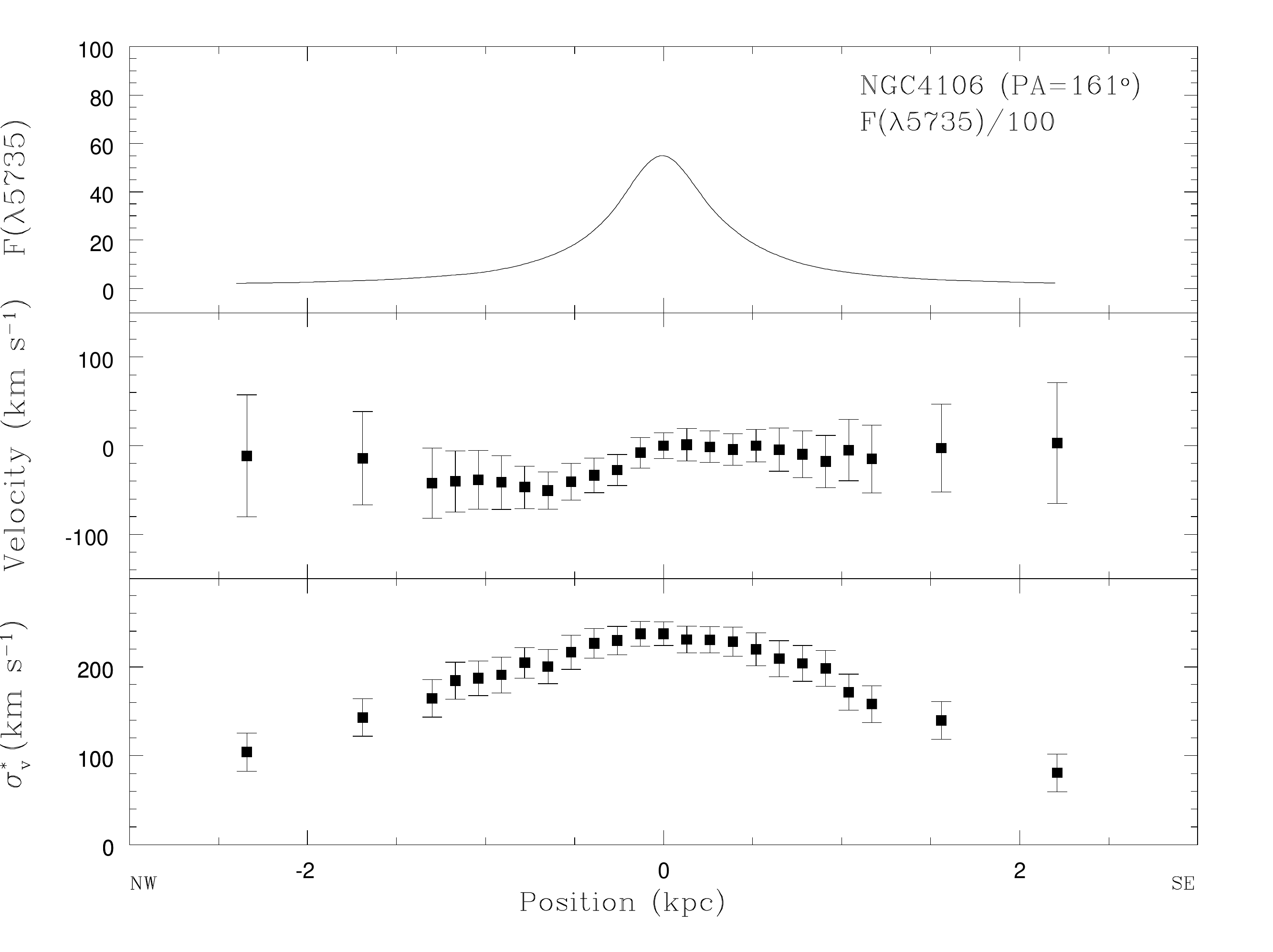}	
	\includegraphics[angle=270,width=0.85\columnwidth]{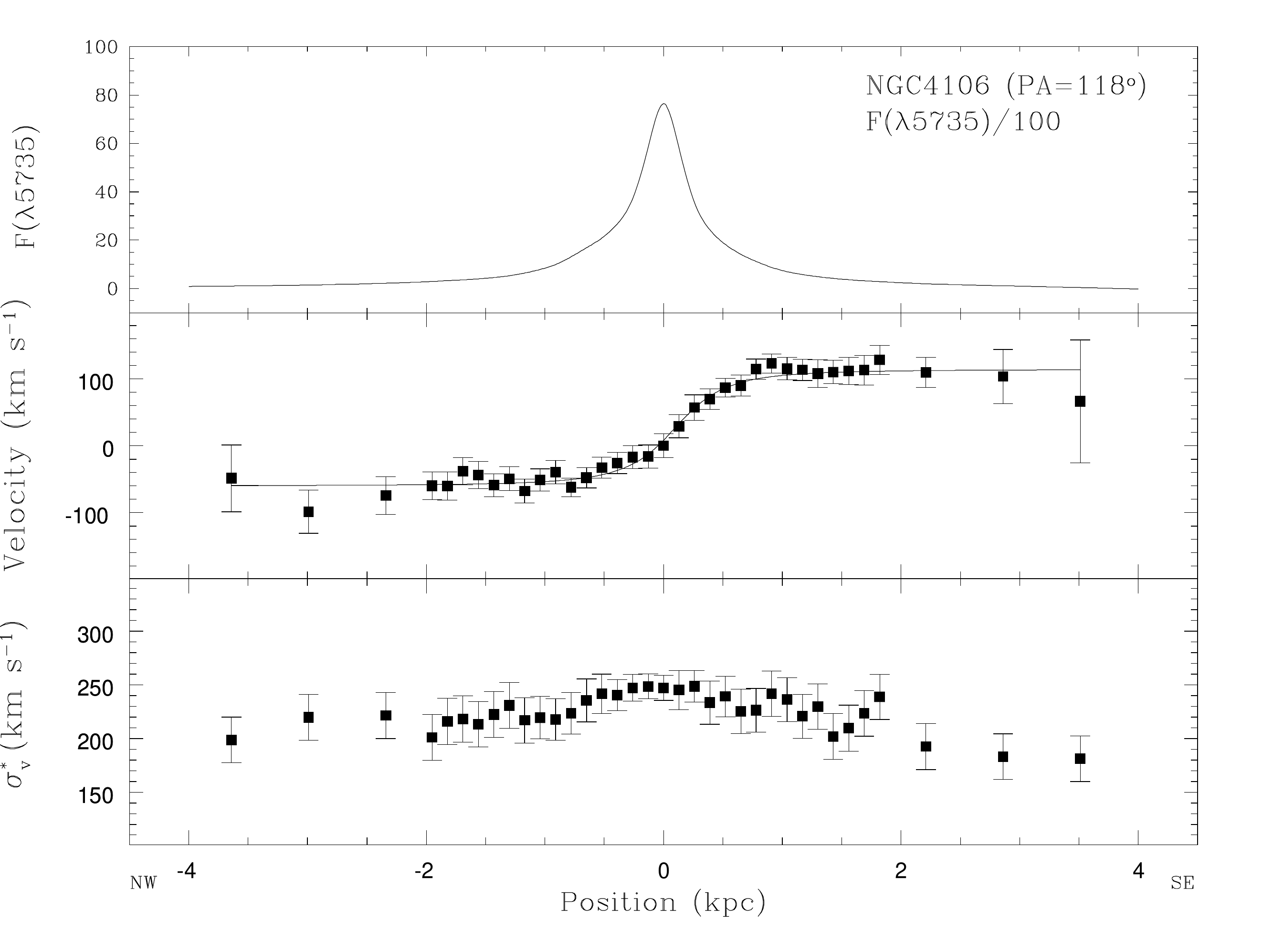}	
	\caption{Spatial profiles of observed $\lambda\,5735$ flux, in units of $10^{-17}$\,ergs\,cm$^{-2}$\,s$^{-1}$ (top), radial velocities after subtraction of the systemic velocities (middle) and velocity dispersion (bottom) for NGC\,4105  and NGC\,4106 along the different position angles, as indicated.The velocity scale was not corrected by the disk inclination.}
\label{fig:cine_A}
\end{figure}

\section{Kinematics}
\label{sec:cine}

\subsection{Stellar Kinematics}
\label{sec:cine_stellar}

The stellar radial velocity and line-of-sight velocity dispersion $\sigma^{*}_{\rm v}$ for each aperture 
were estimated from the cross correlation method of its absorption line spectrum against a set of stellar template spectra.
The radial velocity was reduced to a heliocentric frame of reference $cz_{\rm Helio}$, as usual.
The rotation curve and the $\sigma^{*}_{\rm v}$ radial profile on the slit directions have been estimated 
for each galaxy to characterize the dynamical system. 

We adopted the Radial Velocity Analysis package
of Smithsonian Astrophysical Observatory \citep[{\scriptsize\,RVSAO}][]{1998PASP..110..934K},
from {\scriptsize\,IRAF} package to measure the  radial velocity and velocity dispersion.
Spectra of K giant stars
(collected under the same instrumental setup of the galaxy pair observations)
were used as reference for the cross correlations
since optical spectra of evolved ETGs are dominated by them.
The observed K giants are HD\,120452, HD\,132345, and HD\,136028.
Their heliocentric radial velocities are from \citet{1995A&AS..110..177D} catalogue.
The best stellar template from the cross correlation of nuclear aperture spectrum is assumed
as a single template for all aperture spectra of each galaxy.
The spectral range of cross correlations is $\lambda\lambda$3800\,--\,6800{\AA},
and the Fourier filters are those suggested by \citet{1979AJ.....84.1511T}.
A second order polynomial function is employed to fit half height of the cross correlation peak
for estimating its full width at half maximum  (FWHM).

The observed radial velocity is simply read from the position of correlation peak, and its correspondent error is 
directly given by {(3/8)$\times$(FWHM/(1+$r$))},
where $r$ is the height of the correlation peak divided by the amplitude of a sinusoidal noise for the correlation 
function \citep{1998PASP..110..934K}.
The velocity dispersion of each galaxy aperture spectrum has been estimated
as a function of FWHM through a third order polynomial fit.
To establish the relation between $\sigma_{\rm v}$ and FWHM,
we convoluted all three template spectra took at March 4$^{\rm th}$ night by different Gaussians,
representing $\sigma_{\rm v}$ from 50 up to 500 km\,s$^{-1}$ with steps of 50 km\,s$^{-1}$. Cross correlations 
between the broadened template spectra and the respective observed stellar spectrum provide the $\sigma_{\rm v}$ 
and the FWHM. Errors in velocity dispersions were estimated artificially adding Poisson noise in the broadened spectra 
of a stellar template.

The radial velocities after subtraction of the systemic velocities, the velocity dispersion $\sigma^{*}_{\rm v}$, and the 
spatial profiles of the $\lambda$5735 continuum 
flux across the observed slit positions are shown in Fig.~\ref{fig:cine_A}
(no geometric correction was applied due to the inclination of the galaxy plane relative to the plane of the sky). 
These galaxies are separated by a 1.15\,arcmin in projection, corresponding to 9.6\,kpc and a difference in 
radial velocities of $\sim274$\,km\,s$^{-1}$. These values are very similar to those found by \citet{2012ApJ...753...43K}.

{\bf NGC\,4105.}
No signs of rotation were detected for the stellar component over both slit directions observed for this elliptical galaxy (see  Fig.~\ref{fig:cine_A}).
On the PA$=46.4^{\circ}$ we can observe a U-shaped symmetry in the radial velocity, with values  reaching $209$ and 
$238$\,km\,s$^{-1}$ at $\sim3.36$\,kpc (SW) and $\sim3.24$\,kpc (NE), respectively, and no tendency to decline. 
In Section \ref{sec:simulations}, we present the velocity map of a simulation that describes quite well the dynamics of this pair. We will then come back to this point.
For PA$=118^{\circ}$ the radial velocity and the velocity dispersion are approximately homogeneous along the radius, with a slight decline towards the SE side, reaching a maximum value of
$52$\,km\,s$^{-1}$ at $\sim1.5$\,kpc, probably due to the interaction.
We considered the heliocentric velocity of NGC\,4105 being a radial velocity measured in the continuum peak 
along the slit position at PA$=46.4^{\circ}$, that is, v$_{r}=1949\pm18$\,km\,s$^{-1}$.

\citet{2005MNRAS.363..769S} investigated the dark matter in early-type galaxies from dynamical modelling and found that the velocities are 
not anti-symmetric due to the interaction with the companion. 
These authors also found that the velocity dispersion decreases from the central values of $320-200$\,km\,s$^{-1}$, 
in agreement with our results.

{\bf NGC\,4106. }
This galaxy shows a nearly symmetrical rotation velocity curve across both slit directions.
For the slit position crossing the nucleus, PA$=161^{\circ}$, a heliocentric velocity of $2223\pm15$\,km\,s$^{-1}$ 
and a maximum value of the rotation velocity of $\sim50\pm21$\,km\,s$^{-1}$, was found.
One side of the curve at $\sim0.7$\,kpc towards the southeast of slit position PA$=161^{\circ}$ presents oscillations in the 
flat region of the curve of the order of 30\,km\,s$^{-1}$, where originates the tidal arm.
The most symmetrical, less-scattered, and smooth curve was derived using the slit position at PA$=118^{\circ}$.
Both sides of the curve display regions with fluctuations of about 20\,km\,s$^{-1}$. One $\sim0.5-1.5$\,kpc towards the northwest 
and another at $\sim1-2$\,kpc towards the SE. The maximum rotation velocity is $128\pm22$\,km\,s$^{-1}$ at $\sim1.82$\,kpc. 

The radially symmetric velocity dispersion profile obtained by us along both directions (PA$=161^{\circ}$ and PA$=118^{\circ}$) are in good agreement with the results previously obtained by \citet{1998A&AS..130..267L}. 

Hence, we predict that NGC\,4106 galaxy shows a fairly significant rotational symmetry.
For this reason, we adopted a very simple approximation for the observed velocity distribution. 
Specifically, assuming circular orbits in a plane $P(i,\psi_{0})$ that are characterized  by
its inclination relative to the plane of the sky ($i$) 
and the position angle (PA) of the line of nodes $\psi_{0}$.  It results in an observed radial circular velocity at a 
position $v$(R,$\psi$) in the plane of the sky given by \citet{1991ApJ...373..369B}:

\begin{equation}
\label{eq:ajuste}
 v({\rm R},\psi)={\rm V_{s}}+\tfrac{{\rm V_{0}}{\rm R}cos(\psi-\psi_{0})sin(i)cos(i)}
 {\sqrt{{\rm R^{2}}[sin^{2}(\psi-\psi_{0})+cos^{2}(i)cos^{2}(\psi-\psi_{0})]+{\rm R_{c}^{2}}cos^{2}(i)}},
\end{equation}
where R is the radius in the plane of the galaxy, V$_{s}$ is the systemic velocity, 
and V$_{0}$ and R$_{c}$ are defined as amplitude parameters and shape of the curve.

The inclination of NGC\,4106 with respect to the plane of the sky, estimated as $i\sim50^{\circ}$, was determined by the observed axial ratio taken from \citet{2015A&A...582A..86H}. The position angle of the line of nodes, $\psi_{0}$\,=\,91$^{\circ}$, was also taken from the same reference.  
From the fitting of the radial velocity observed for NGC\,4106 (PA$=118^{\circ}$),  
shown in the bottom panel of Fig.~\ref{fig:cine_A}, we obtain  
a deprojected velocity amplitude of $V_0=138\pm6$\,km\,s$^{-1}$.
Supposing that the mass in a certain radius is given by $M(R)=RV_{0}^{2}/G$, 
using the amplitude of this fit, the dynamical mass of the S0 galaxy is about  $1.6\times10^{10}\,M_{\odot}$ up to a radius of 3.64\,kpc.

\begin{figure}
\centering
  \includegraphics[angle=270,width=0.85\columnwidth]{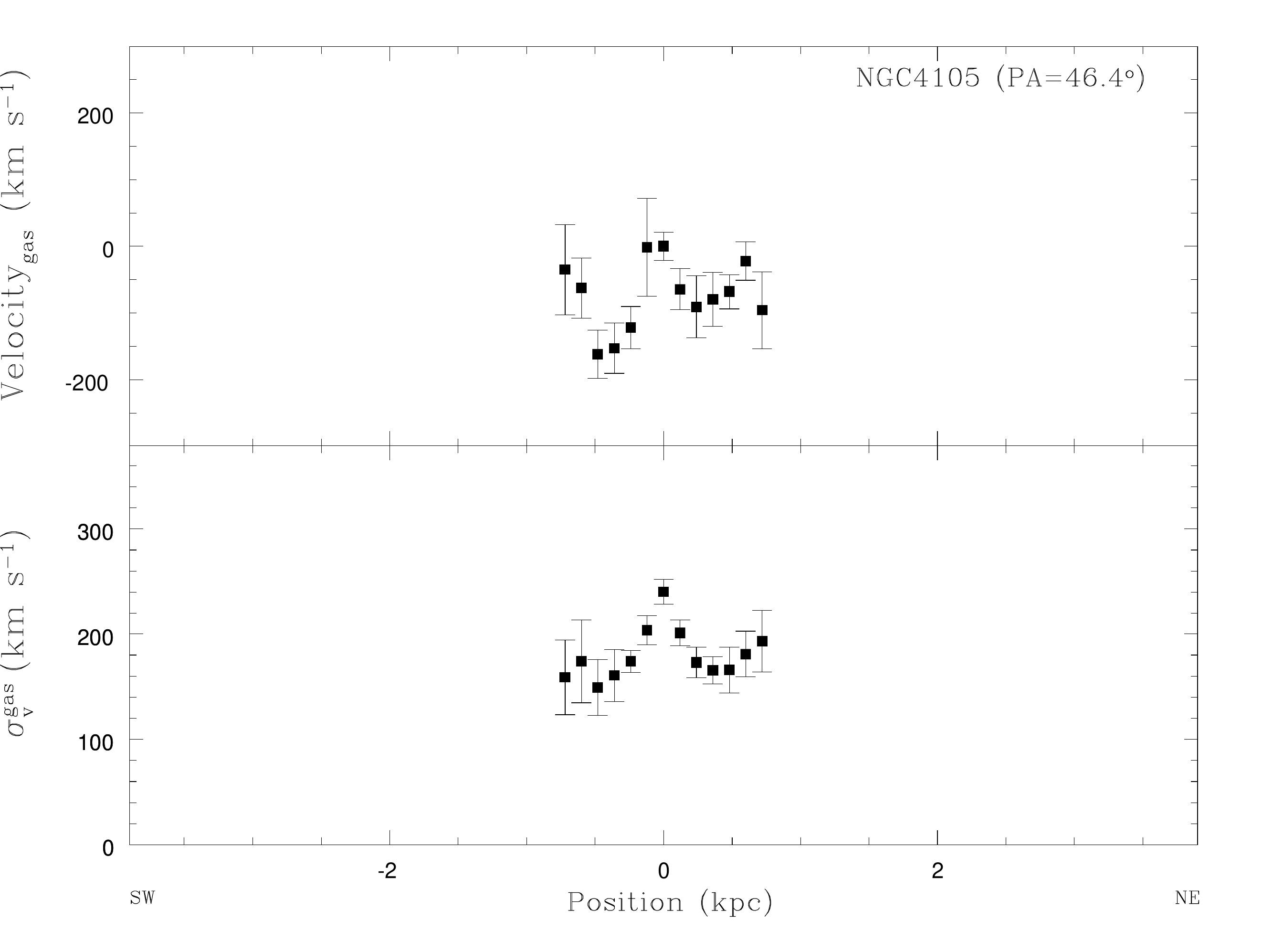}
  \includegraphics[angle=270,width=0.85\columnwidth]{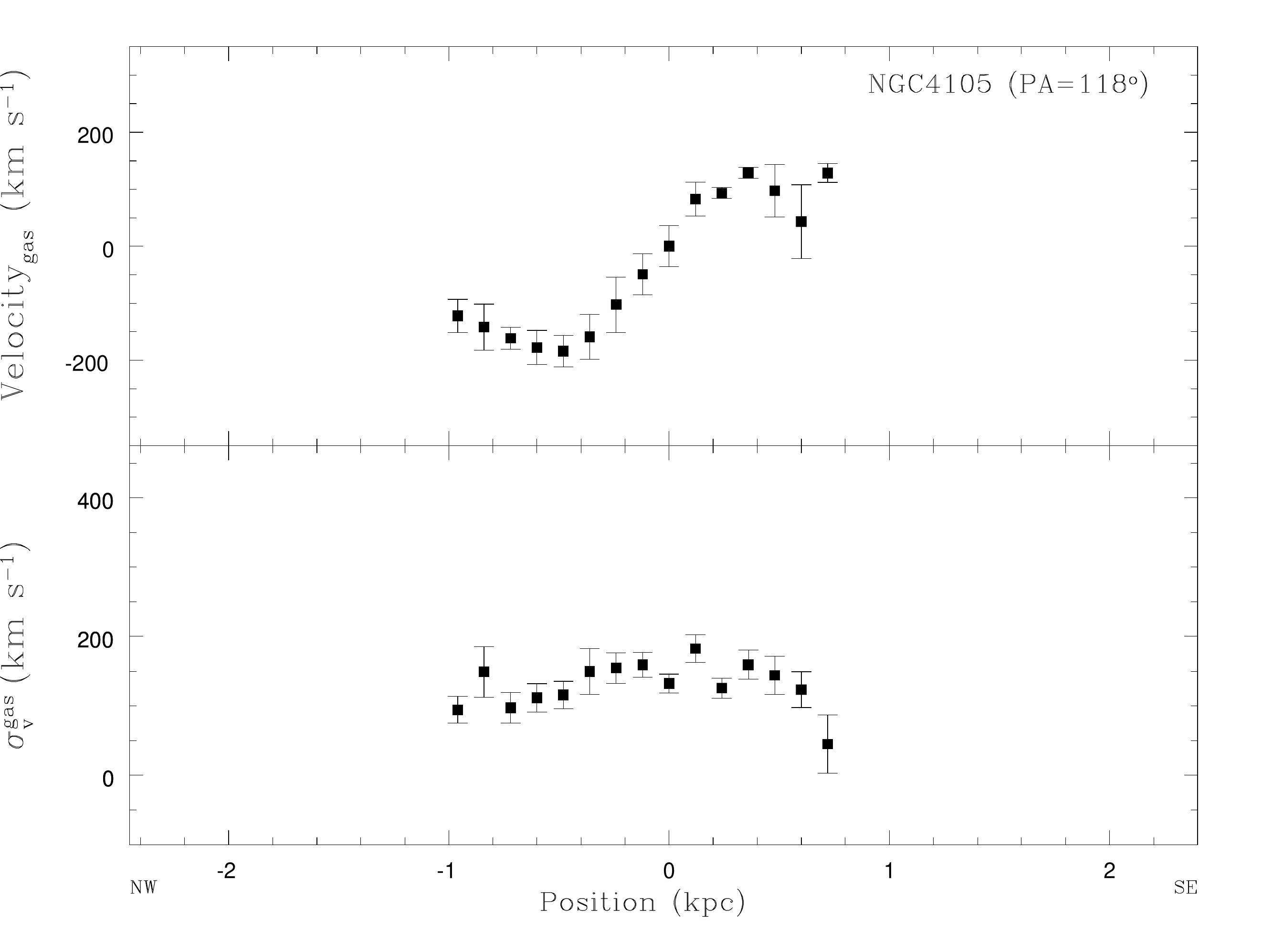}
 \includegraphics[angle=270,width=0.85\columnwidth]{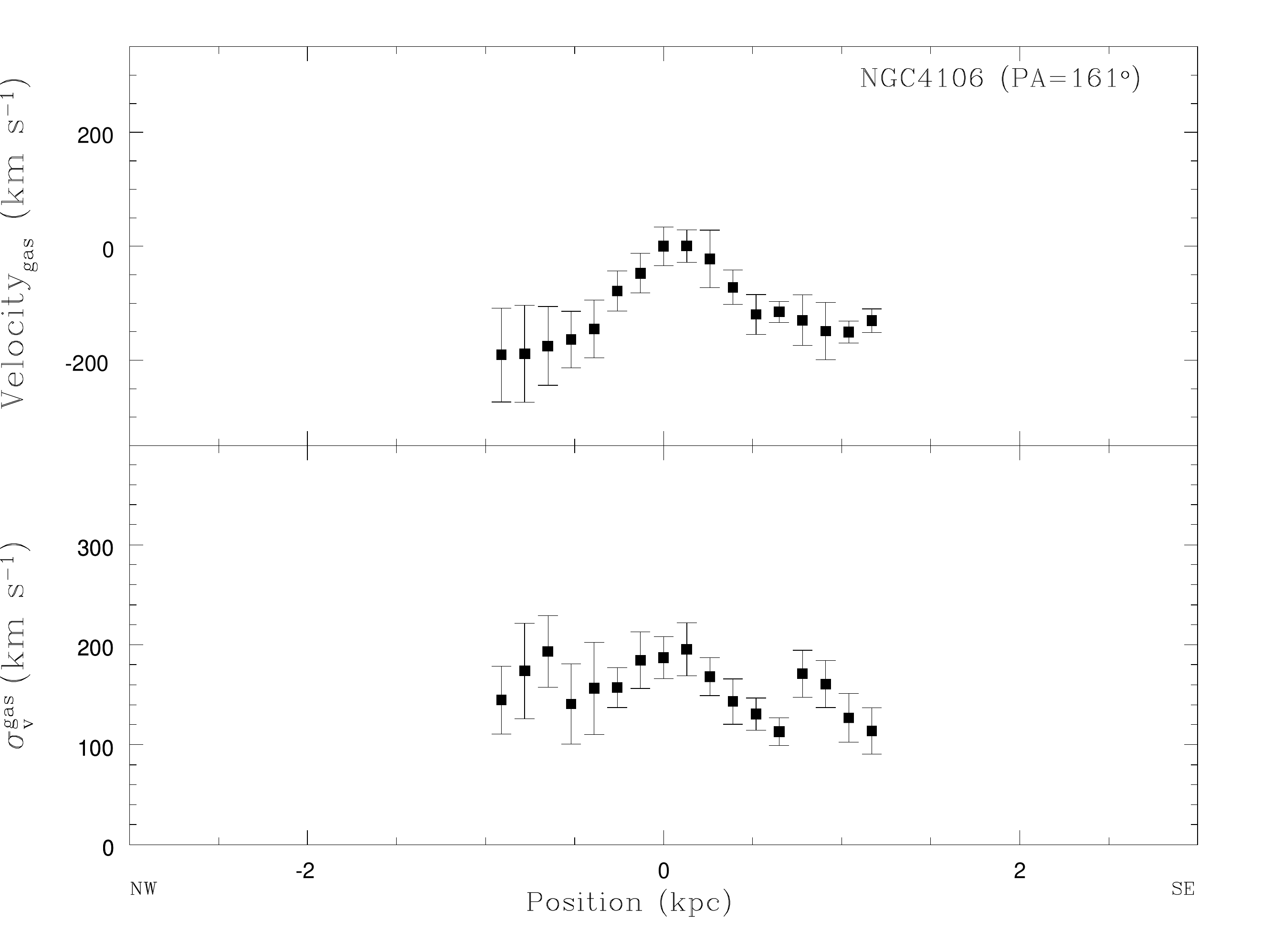}
 \includegraphics[angle=270,width=0.85\columnwidth]{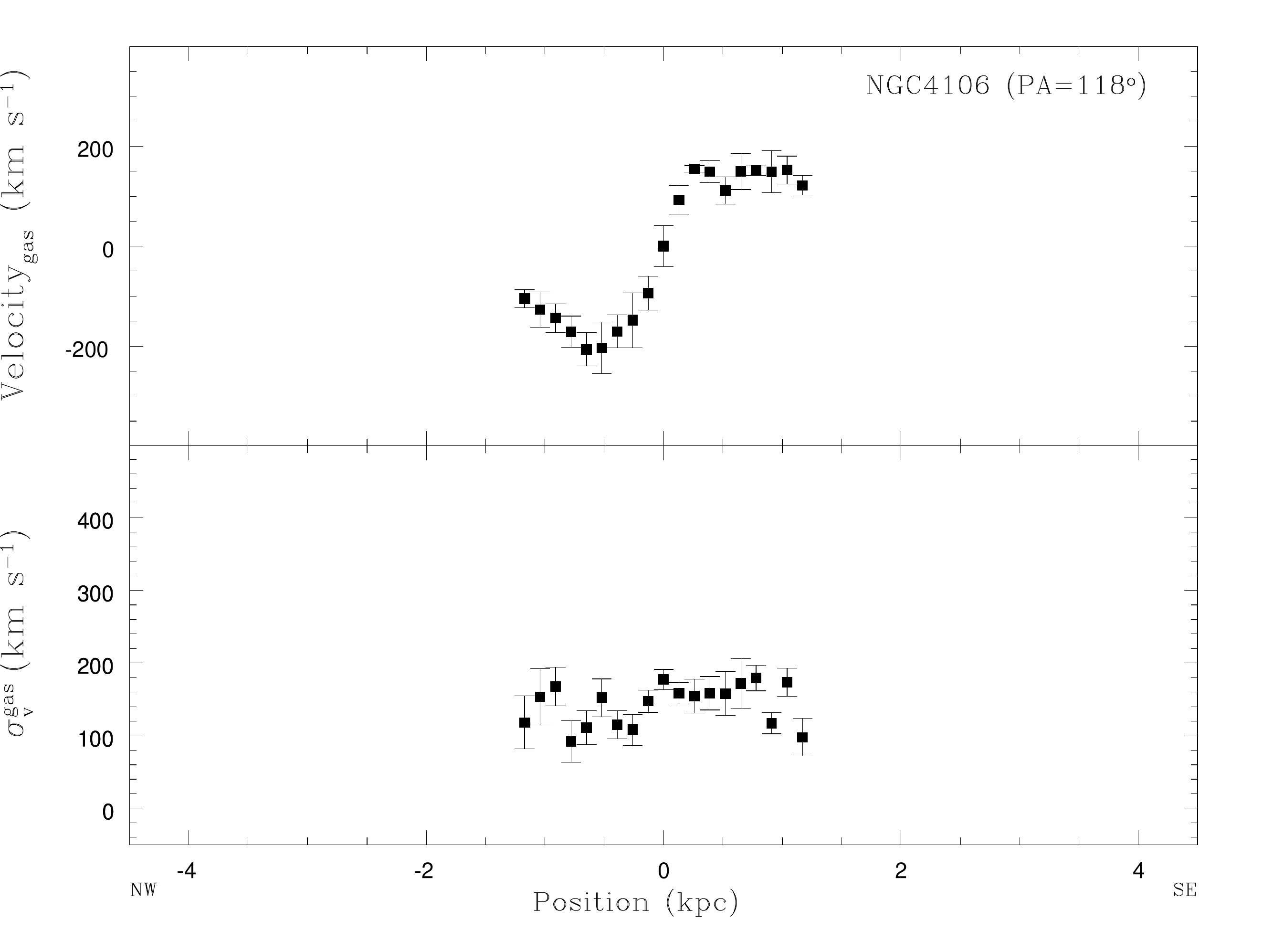}
\caption{Radial velocity and velocity dispersion profiles for the ionized gas in NGC\,4105 and NGC\,4106 derived from [O{\scriptsize\,II}]3727\AA{} and [N{\scriptsize\,II}]6583.5\AA{}. Radial velocities were obtained after subtraction of the velocity in the central aperture, defined by the position of the flux peak in the spatial  profile. 
As in Fig. \ref{fig:cine_A}, the velocity scale was not corrected for galaxy inclination.}
    \label{fig:gas_cine}
\end{figure}

\subsection{Gaseous Kinematics}
\label{sec:gas}

The gaseous radial velocity in NGC\,4105 and 
NGC\,4106  was determined using the task {\scriptsize\,EMSAO} 
\citep{1995ASPC...77..496M}
implemented in the {\scriptsize\,RVSAO} package \citep{1998PASP..110..934K}. This program finds emission lines in a spectrum and computes the 
observed 
centre, yielding individual radial velocity and error for each line and combining  them into a single radial velocity. The program was successful at finding the emission lines [O{\scriptsize\,II}]3727\AA{} and [N{\scriptsize\,II}]6584\AA{}. The errors are estimated from the uncertainty in the dispersion function in Angstroms and in  fit to the center of the Gaussian. The velocity dispersion $\sigma^{\rm gas}_{\rm v}$ and their error were calculated as a function of FWHM, measured using the {\scriptsize\,SPLOT} task from {\scriptsize\,IRAF}. The FHWM uncertainty is estimated from the Poisson error. The velocity dispersion was corrected by the instrumental broadening $\sigma_{\rm v,inst}$ ($132\pm5\,{\rm km\,s}^{-1}$),  
where $\sigma_{\rm line}$ $\sim$\,($c$/2.352)\,$\times\,{\rm FWHM}_{\rm line}$/$\lambda_{\rm central}$
and $\sigma_{\rm line}^{2}$\,=\,$\sigma_{\rm v}^{2}$\,+\,$\sigma_{\rm v,inst}^{2}$. We assumed the radial velocity of the 
central aperture of  each galaxy as the heliocentric velocity. 

The radial velocity and velocity dispersion  profiles of each observed slit positions are shown in Fig.~\ref{fig:gas_cine}.
The number of kinematic measurements of the gaseous component is smaller than the stellar one, reaching only the central kpc of galaxies.

As can be seen in the top panel of Fig.~\ref{fig:gas_cine}, NGC\,4105  does not show a U-shaped radial velocity profile along PA\,=\,46.4$^{\circ}$, like that observed for the  
stellar kinematics. Gas velocity dispersions on this slit are lower than that of the stars, but the profile is similar, with a peak in the nuclear region and descending wings for both sides. The radial velocity profile across PA\,=\,118$^{\circ}$ shows a clear rotational pattern, very different from the almost flat profile of the stellar component, considering our error bars (see Fig. \ref{fig:cine_A}). It should be noted that the slit did not cross the center of the galaxy. The behavior of ionized gas velocity profiles in both directions is very different from that of the stellar profiles, i.e., ionized gas and stars do not share the same dynamics.

The gaseous  and  stellar  kinematics in NGC\,4105 was previously studied by \citet{2000ApJS..127...39C}, using long-slit spectroscopy and a similar methodology to ours. They observed with the slit positioned at PA\,=\,156$^{\circ}$, different than our slit position and also not coincident with the photometric major axis position angle (see Tab.\ref{tab:sample}, PA\,=\,136.4$^{\circ}$). \citet{2000ApJS..127...39C} also  verified that the gas velocity curve shows a strong rotation pattern with peak velocities above 200\,km\,s$^{-1}$. Their stellar velocities show a quite symmetric profile with positive velocities towards the NW direction and negative to the SE ($\leq 40$\,km\,s$^{-1}$ for both sides), which they interpret as rotation.  Based on it, they argue that the ionized gas in NGC\,4105 is counter-rotating with respect to the stars, is of external origin, is not in equilibrium, and was acquired recently, also affecting stellar kinematics. In the stellar velocity profile of our PA$=118^{\circ}$ slit position, we also observed counter-rotation (with a slightly lower velocity range). But our simulation fails to reproduce that. We show in Section \ref{sec:simulations} that the origin of this gas is not the current companion galaxy, as can be inferred from our simulation, in which no gas is transferred from NGC\,4106 to NGC\,4105.
 
We argue that the stellar velocity profile obtained by \citet{2000ApJS..127...39C} may reflect the transfer of orbital angular momentum to the outermost stars of NGC\,4105 during the close encounter, in contrast to a pre-existing rotation.

In NGC\,4106, the radially symmetric velocity profile across the slits  reach a maximum velocity of $206\pm33$\,km s$^{-1}$ at $\sim-0.65$\,kpc over PA$=118^{\circ}$. The velocity dispersion profiles at both slits are very steeper than that of the stellar profiles, 
and nearly symmetrical up to $\sim$1.2 kpc, with a central value  of $187\pm21$\,km s$^{-1}$ for PA$=161^{\circ}$ and $178\pm14$\,km s$^{-1}$ 
in the central aperture of PA$=118^{\circ}$.

The H{\scriptsize I} gas masses for the galaxies were measured by \citet{1979A&A....74..172B}, who  
found $M_{\textrm H{\scriptsize\textrm I}} = 5.2\times10^{8}$\,M$_{\odot}$ for 
NGC\,4105 and 
$M_{\textrm H{\scriptsize I}} = 3.4\times10^{8}$\,M$_{\odot}$ for NGC\,4106.

\section{N-Body + SPH simulations}
\label{sec:simulations} 
Numerical simulations were carried out in order to reconstruct the interaction history between the galaxies NGC\,4105 and NGC\,4106. 
The simulations were performed using the TreePM/SPH code {\sc P-GADGET3}, an updated and significantly extended version of 
{\sc GADGET-2} \citep{2005MNRAS.364.1105S}, where the dark matter and stars are followed by collision less particles and gravity 
is calculated in a tree-code. The gas is treated through Smoothed Particle Hydrodynamics  \citep[SPH,][]{2002MNRAS.333..649S}, 
with star formation process, radiative cooling and feedback, and sub-grid star formation model as defined in \cite{2003MNRAS.339..289S}, 
from which we used the same parameters.

\subsection{Initial conditions for the simulation}\label{sec:initcond}

Based on the observational data presented here and elsewhere, we began to construct the initial conditions for our simulations. 
In the case of  AM\,1204-292, the collision is between an elliptical galaxy (NGC\,4105) and a S0 galaxy (NGC\,4106). 

The total stellar mass of both galaxies were  derived through the mass-to-light ratios resulting from stellar population synthesis (see Sect. \ref{sec:sps}). We correct the slit stellar mass obtained from {\scriptsize\,STARLIGHT} to total stellar mass by scaling the total and slit luminosity in the B-band. The total B-band luminosity was taken from the ESO-LV catalogue and are 
L$_{B}\sim 7.24 \times\,10^{10}$\,L$_{\odot}$ and 
L$_{B}\sim5.62\times\,10^{10}$\,L$_{\odot}$ for NGC\,4105 and NGC\,4106, respectively. Then, the stellar masses are $1.81\times\,10^{11}\,M_{\odot}$ and $8.30\times\,10^{10}\,M_{\odot}$, for NGC\,4105 and NGC\,4106 respectively.

\begin{table}
\begin{small}
\caption{NGC\,4105 (E with gas) model parameters.}
\begin{tabular}{l r }
\hline \hline \smallskip
 & Phys. Units$^{*}$  \\
\hline \noalign{\smallskip}
\multicolumn{2}{l}{\bf NGC\,4105 model (Elliptical)}   \\
Number of particles in halo & 700,000  \\
$R_{200}$  & 155 kpc   \\
$M_{200}$  & $86.59\times 10^{10}\,\mathrm{M}_\odot$     \\
$V_{200}$  & 155\,km\,s$^{-1}$   \\
$r_h$  & 23.4\,kpc   \\
Halo mass $(M_{\rm{dm}})$ & $71.69\times 10^{10}\,\mathrm{M}_\odot$   \\
&  \\
Number of particles in stellar disk & 200,000  \\
Disk mass $(M_{\rm{d,s}})$ &  $0.086\times10^{10} \,\mathrm{M}_\odot$ \\
Disk radial scale length $(R_{\rm{d,s}})$ &  3.2\,kpc \\
Disk vertical scale thickness $(z_{\rm{0,s}})$ &  0.63\,kpc    \\
 &  \\
Number of particles in gas disk & 200,000 \\
Gas disk mass $(M_{\rm{d,g}})$ &   $0.064 \times10^{10} \,\mathrm{M}_\odot$ \\
Gas disk radial scale length $(R_{\rm{d,g}})$  & 3.2\,kpc \\
Gas disk vertical scale thickness $(z_{\rm{0,g}})$ &  0.035\,kpc \\
 &  \\
Number of particles in spherical stellar component & 300,000  \\
Spherical component mass $(M)$ &  $14.72\times10^{10} \,\mathrm{M}_\odot$ \\
Spherical component radial scale length $(r_h)$ &  4.77\,kpc \\
&  \\
Total mass of the model & $86.59\times 10^{10}\,\mathrm{M}_\odot$    \\
&  \\
Total barionic mass of the model & $15.70\times 10^{10}\,\mathrm{M}_\odot$    \\
\hline
\label{tab:model_par_4105}
\end{tabular}

\end{small}
\end{table}

\subsection{NGC\,4105 model}    

The elliptical galaxy was modeled by two spheroidal  \citet{1990ApJ...356..359H} components, one for the stars, other for the dark matter halo:

\begin{equation}
    \rho(r)=\dfrac{M}{2\pi}\dfrac{r_h}{r(r+r_h)^3}\label{eq:halo_bulge}
\end{equation}

\noindent {where $M$ is the total mass and $a$ is the radial scale length of each component. Two models have were constructed: (i) The first one has a spherical stellar component within a concentric spherical dark matter halo to test the hypothesis that the gas present in this galaxy was stolen from NGC\,4106 during the collision, as will be discussed in section \ref{sec:orbit}; (ii) Stellar and gaseous disks were added to the second model, with exponential density profiles:}

\begin{equation}
    \rho(d)=\dfrac{M_d}{4\pi R_d^2z_0} \exp(-\dfrac{R}{R_d})\rm{sech}^2\left(\dfrac{z}{z_0}\right),\label{eq:disk}
\end{equation}

\noindent {where $M_d$, $R_d$ and $z_o$ are the stellar or gaseous disk
mass, radial scale length and vertical scale length, respectively. }

We start setting up our galaxy model by the values of the mass. The stellar mass is presented in the previous section. For the halo model, we took typical parameters for elliptical galaxies. The first guess for the total mass (dark plus barionic) was obtained using equation 4.249b from \citet{2008gady.book.....B}, based on the scalar Virial theorem. The velocity dispersion is presented in Section \ref{sec:cine_stellar} (Fig. \ref{fig:cine_A}), from the PA\,$=46.4^\circ$, which gave us a mass of $M\sim5\times10^{11}$M$_\odot$. The radial velocity curves are also used as observational beacons. During the construction of the model galaxy, we should find a good set of parameters that are committed to observational data, and still produce a dynamically stable model. The final value of the total mass is a little higher: $M_{200}\sim8.7\times10^{11}$M$_\odot$, as the stellar mass. 

All galaxy models in the present paper were set following the prescription given by \citet{2005MNRAS.364.1105S}. Spheroidal halo and bulge follow the \citet{1990ApJ...356..359H} profile (eq. \ref{eq:halo_bulge}), while the gas and stellar disks have an exponential density profile (eq. \ref{eq:disk}).  The parameters of the final model are given in Table~\ref{tab:model_par_4105}. 

\subsection{NGC\,4106 model}    

For NGC\,4106, the S0 type galaxy, we started with the stellar mass derived from the M/L ratio obtained from the stellar population synthesis (see Sect. \ref{sec:initcond}, and \ref{sec:sps}. For each model generated, we compare the rotation curve with the stellar radial velocity curves obtained from longslit spectroscopy (Fig. \ref{fig:cine_A}). The slit position PA\,$=161^\circ$ crosses the galaxy through its nucleus, in a direction close to the galaxy minor axis (at PA\,$=181^\circ$), so that no rotation is detected. The other one, at PA\,$=118^\circ$, is close to the major axis direction (at PA\,$=91^\circ$), but does not cross the galactic center. Both radial velocity curves were corrected to estimate the rotation curve of the galaxy, as explained in Section~\ref{sec:cine_stellar}.  Then we generated a set of S0-like galaxy models, constructed with a bulge and a halo given by \citet{1990ApJ...356..359H} spheroids (Eq. \ref{eq:halo_bulge}) and an exponential disk (Eq. \ref{eq:disk}). We 
plotted the corresponding rotation curve over the observed one and choose the one that best reproduces the mean observed rotation curve and is dynamically stable (i.e., it survives with its mass distribution nearly unaltered in a 1\,Gyr run). It was not a fit procedure. The parameters of the final NGC\,4106 model are also given in Table \ref{tab:model_par_4106}.

\begin{table}
\begin{small}
\caption{NGC\,4106 (S0) model parameters.}
\begin{tabular}{l r }
\hline \hline \smallskip
 & Phys. Units$^{*}$  \\
\hline \noalign{\smallskip}
\multicolumn{2}{l}{\bf NGC\,4106 model (S0)}   \\
Number of particles in halo & 500,000  \\
$R_{200}$  & 160 kpc   \\
$M_{200}$  & $95.24\times 10^{10}\,\mathrm{M}_\odot$     \\
$V_{200}$  & 160\,km\,s$^{-1}$   \\
$r_h$  & 32.35\,kpc   \\
Halo mass $(M_{\rm{dm}})$ & $90.94\times 10^{10}\,\mathrm{M}_\odot$   \\
 &  \\
Number of particles in stellar disk & 450,000  \\
Disk mass $(M_{\rm{d,s}})$ &  $3.8 \times10^{10} \,\mathrm{M}_\odot$ \\
Disk radial scale length $(R_{\rm{d,s}})$ &  2.8\,kpc \\
Disk vertical scale thickness $(z_{\rm{0,s}})$ &  0.56\,kpc    \\
 &  \\
Number of particles in gas disk & 200,000 \\
Gas disk mass $(M_{\rm{d,g}})$ &   $0.064 \times10^{10} \,\mathrm{M}_\odot$ \\
Gas disk radial scale length $(R_{\rm{d,g}})$  & 2.8\,kpc \\
Gas disk vertical scale thickness $(z_{\rm{0,g}})$ &  0.56\,kpc \\
 &  \\
Number of particles in bulge & 50,000  \\
Bulge mass $(M)$ &  $0.48\times10^{10} \,\mathrm{M}_\odot$ \\
Bulge radial scale length $(r_h)$ &  0.56\,kpc \\
&  \\
Total mass of the model & $95.52\times 10^{10}\,\mathrm{M}_\odot$    \\
&  \\
Total barionic mass of the model & $4.35\times 10^{10}\,\mathrm{M}_\odot$    \\
\hline
\label{tab:model_par_4106}
\end{tabular}
\end{small}
\end{table}

Each galaxy model was evolved in isolation for 3\,Gyr to test stability and to allow its numerical relaxation.  

\begin{table}
\begin{small}
\caption{Orbital parameters from the best model. }
\begin{tabular}{l r }
\hline \hline \smallskip
 & Phys. Units$^{*}$  \\
\hline \noalign{\smallskip}
\multicolumn{2}{l}{\bf NGC\,4106 model (S0)}   \\
Eccentricity  & 0.9 \\
Pericenter distance  &   5.7\,kpc \\
Present distance  &  17.7\,kpc \\
$V_{sys}$ & -329\,km\,s$^{-1}$ \\
Pericenter vector direction$^{*}$  & (0.004, 0.961, -0.275)   \\
Orbital plane normal vector direction$^{**}$ &  (0.356,  -0.871,  -0.338) \\
Present day position vector$^{*}$ & (-8.8, -4.2, -14.8)\,kpc \\
Present day velocity vector$^{*}$ & (-792, -91, -329)\,km\,s$^{-1}$ \\
Initial galaxy separation  & 1.0\,Mpc \\
\hline
\label{tab:model_par}
\end{tabular}
{\bf Notes:} 

$^{*}$Vector components defined in a Cartesian reference
frame (X, Y, Z) centered on NGC\,4105,  with X to the North, Y to the West, and Z in the line
of sight direction, with positive values towards the observer.

$^{**}$The orbital plane normal vector is the direction of the orbital
angular momentum (reference frame defined as in $^{*}$).

\end{small}
\end{table}

\begin{figure*}
    \begin{center}
    \hspace{-1.4mm}
    	\includegraphics[width=\columnwidth]{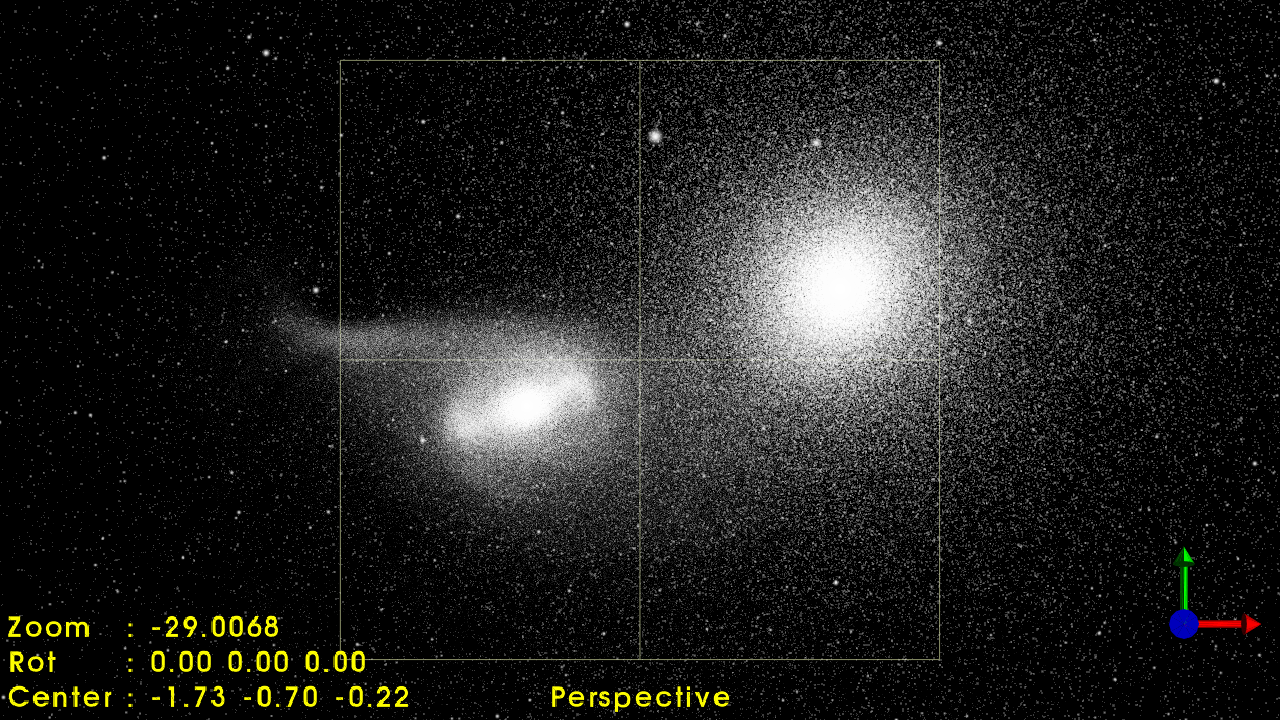}
    	\includegraphics[width=\columnwidth]{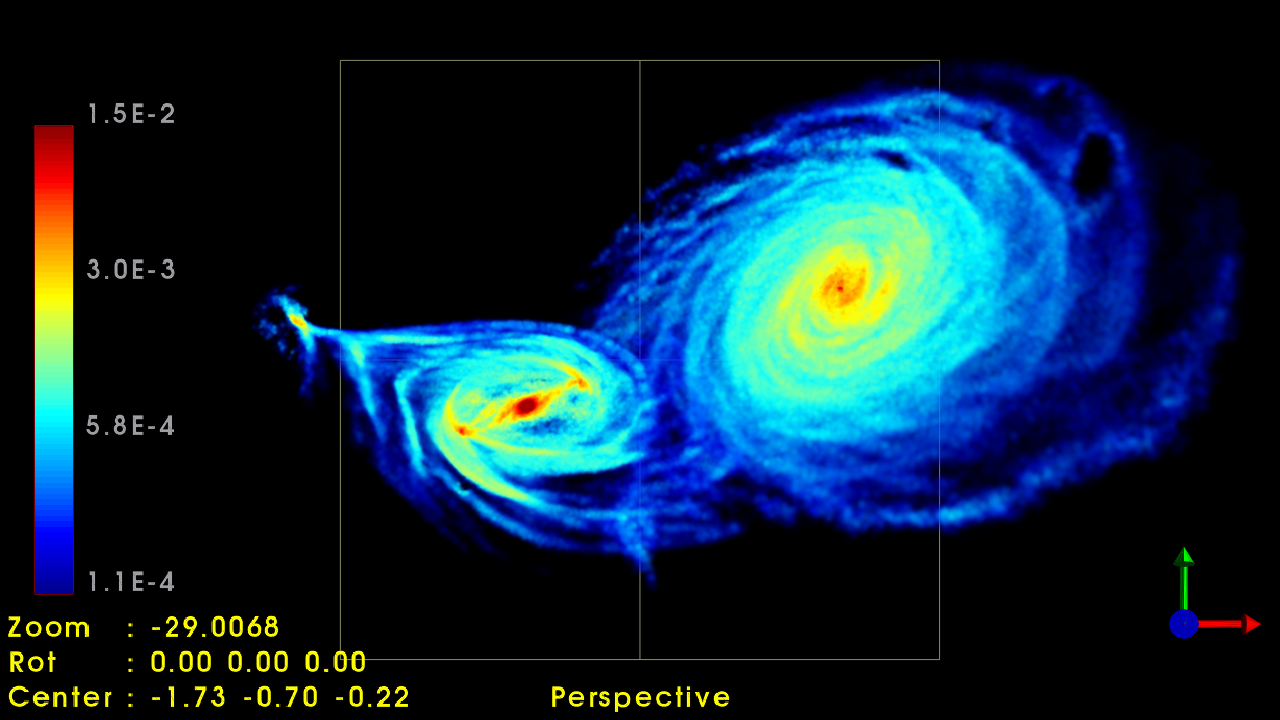}	
    	\includegraphics[width=\columnwidth]{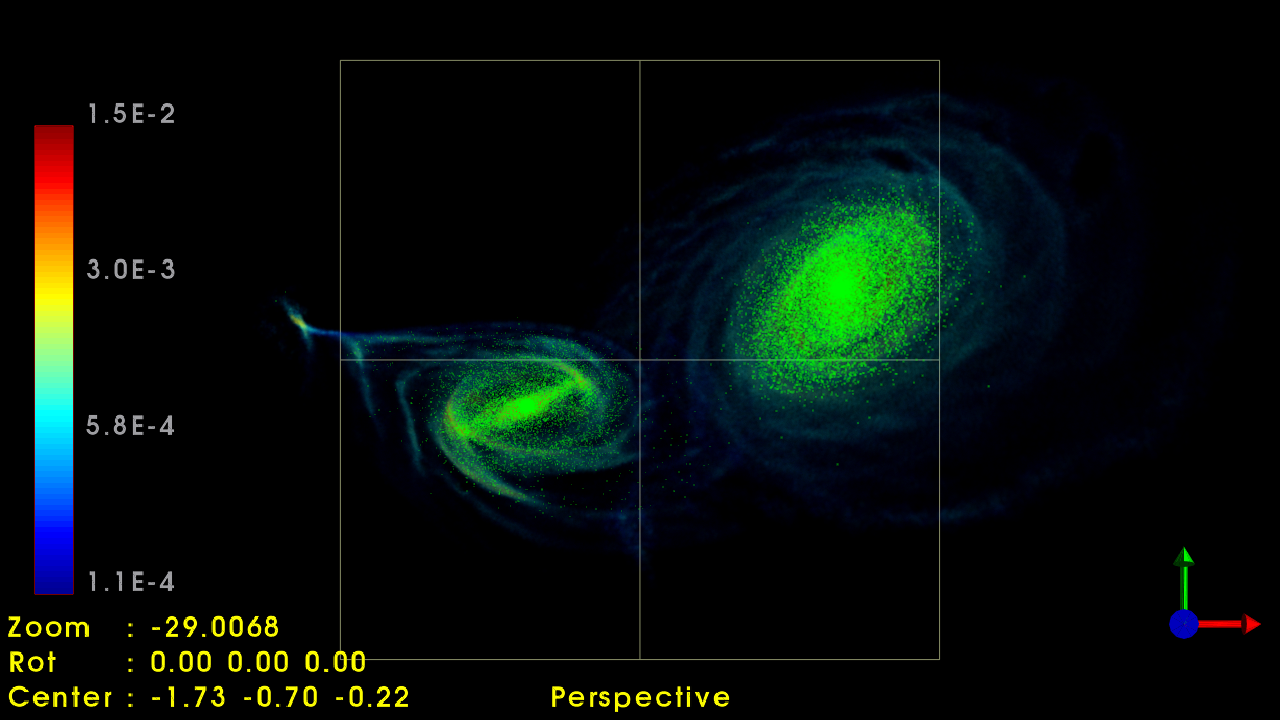}
    	\includegraphics[width=\columnwidth]{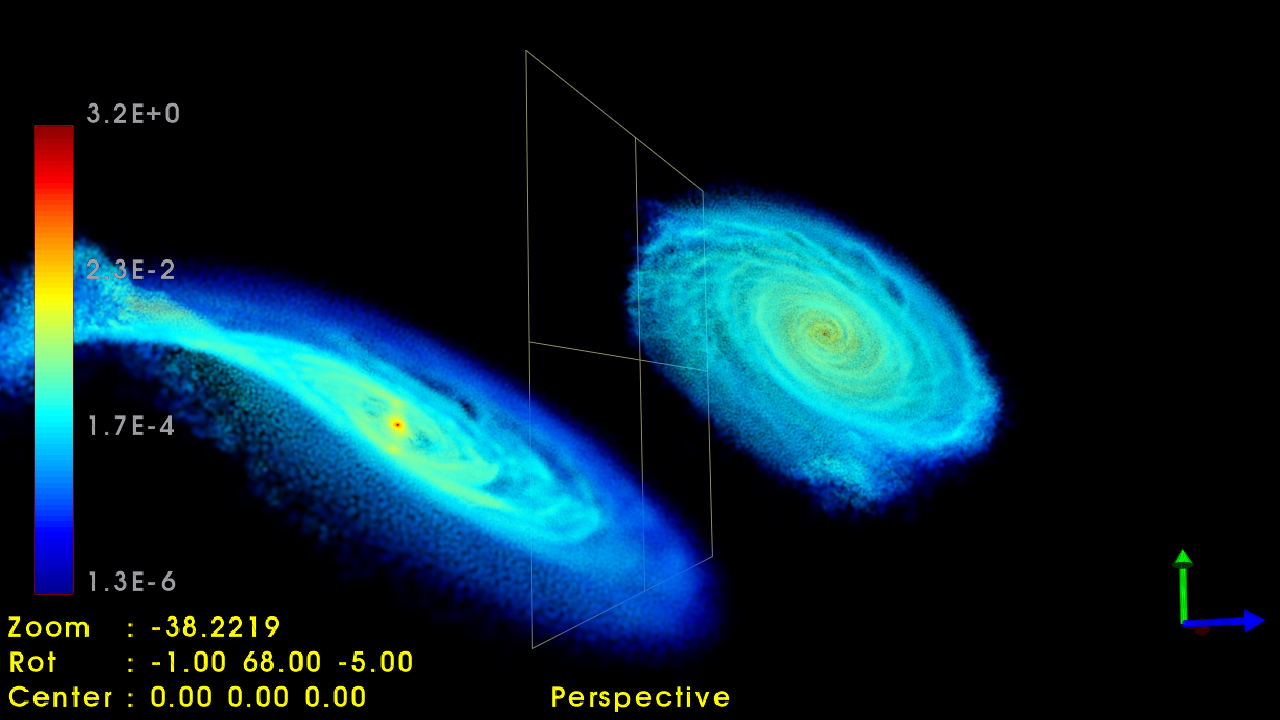}
        \caption{Selected views of the snapshot that best reproduces the observed kinematics and morphology of AM\,1204-292, 14.2 Myr after perigalacticum. Top left panel: Only stellar particles are shown. Top wright: Gas density distribution. Bottom left panel: stars formed during the simulation are drawn in green over the gas distribution. Bottom right panel: The gas distribution vied from a different view point to show that the galaxies are completely detached from each other, with no connecting bridges. No gas is transferred between galaxies. The square grid at the center of each panel is 20\,kpc wide.}
    \label{fig:simulations}
    \end{center}
\end{figure*}    

\subsection{The orbit} \label{sec:orbit}

In order to find the orbit followed by both galaxies, we wrote a code that takes as input parameters the distance to the galaxy pair, the positions of the galaxies projected in the sky plane, the radial velocities of their mass centers. Then the algorithm calculates orbits that may lead to the present state of the system. Many thousands of orbits are presented. Then we identify families of orbits and select those that, by morphological and kinematic constraints, are suitable to take the system to its present state. Then, a large series of low resolution test simulations were run, out of which, 6 best cases were chosen and ran in higher resolution. Orbital parameters are presented in Table~\ref{tab:model_par}.

The simulation starts 1\,Gyr before perigalacticum, with the galaxies at an initial distance of 1\,Mpc. In this initial configuration, dark matter halos are far enough  from each other, so the dynamical damage caused by the sudden presence of the other galaxy model is negligible. Galaxy models were evolved in isolation for 1\,Gyr, before being stacked together for the collision run. 

A whole set of simulations was performed using a simple elliptical galaxy model for NGC\,4105 (spherical model with dark matter and stars), with the gaseous content of the pair entirely placed in the NGC\,4106 model. Remember that both galaxies have a similar amount of H{\scriptsize\,I} gas, as stated at the end of Section~\ref{sec:gas}.  Despite being very successful in several aspects (reproduction of morphology, kinematics, consistent evolution of the star formation rate, etc ...), the model failed in one respect: there was no gas transfer from NGC\,4106 to NGC\,4105. From this result, we can be sure that, as already suggested by \citet{2000ApJS..127...39C}, the gas present in NGC\,4105 is pre-existing, and must have been captured in a previous event.

\begin{figure*}
    \centering
	    \includegraphics[angle=0,width=18cm]{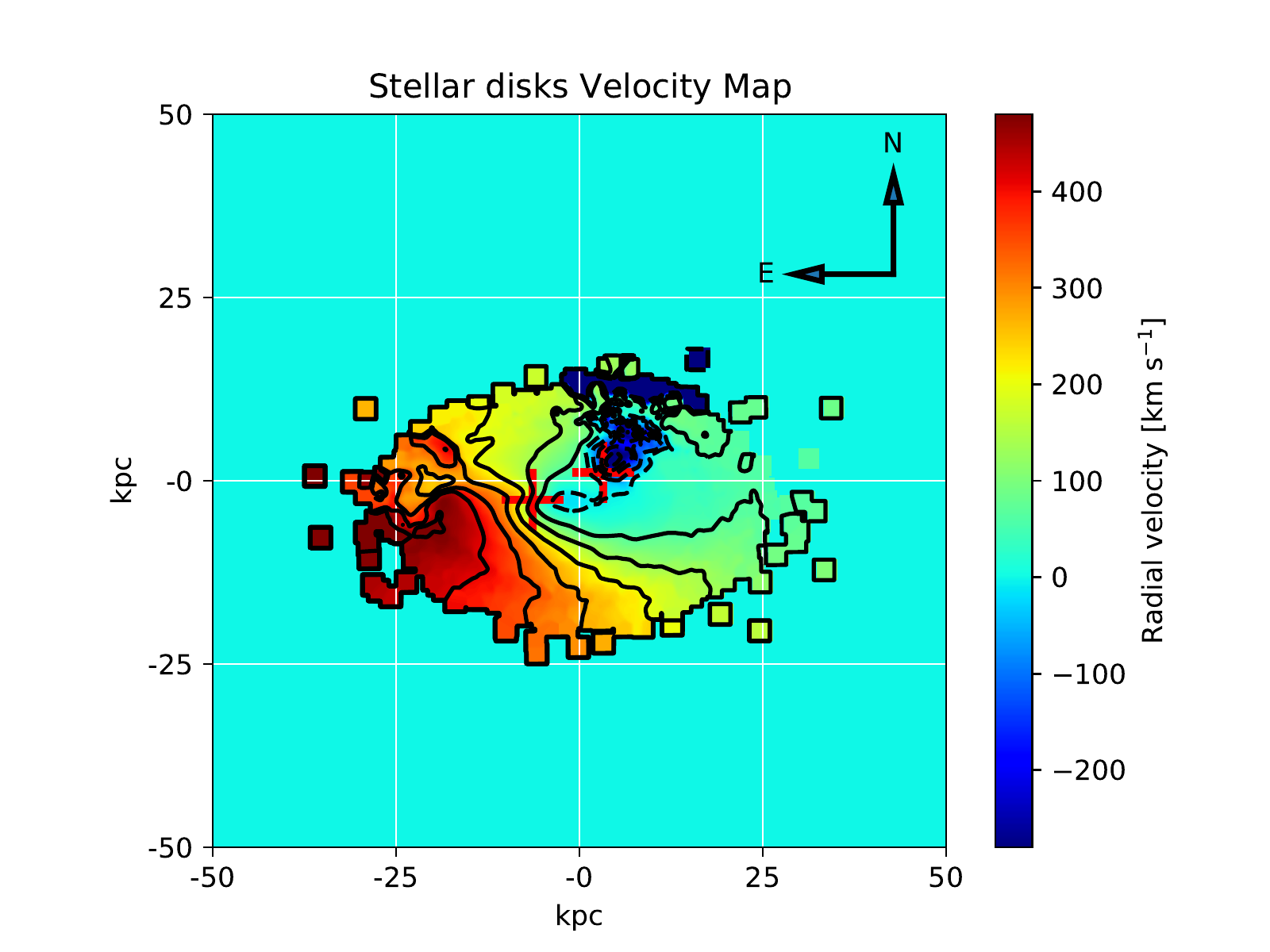}
        \caption{ Stellar radial velocity map of the snapshot that best reproduces the observed kinematics and morphology of AM 1204-292, 14.2 Myr after perigalacticum. Only disk components of both galaxies are shown. Red crosses mark the center of each model galaxy. }
        \label{fig:velmap_sec}
\end{figure*}

Then, the second set of simulations was run with stellar and  gaseous disks placed added to the NGC\,4105 model. It makes sense in view of the results by \cite{2009A&A...502..473G}, who found dusty features  in  the  central region and modeled the brightness distribution, concluding that it presents a disky outer  structure. The subtraction of their model showed a central bar-like structure. 
From the best fit simulation out of this set, we present in Fig.~\ref{fig:simulations} different renderings of the snapshots that reproduces the current situation of the system, 14.2\,Myr after perigalacticum. The main morphological features, such as the general shape and the tidal arms, are well reproduced by the simulation. In the last panel of Fig.~\ref{fig:simulations} we show the gas disks from a viewpoint different than our line of sight. The disks are completely disconnected, with no bridges, connecting tails or stream between them, showing that no gas is transferred from one galaxy to the other.

The velocity distribution, measured by long slit spectroscopy, can also be compared directly to the velocity map extracted from the simulation at the current system state. It is presented in Fig.~\ref{fig:velmap_sec}, where only the stellar disks are shown.

\begin{figure}
\centering
    \includegraphics[angle=0,width=\columnwidth]{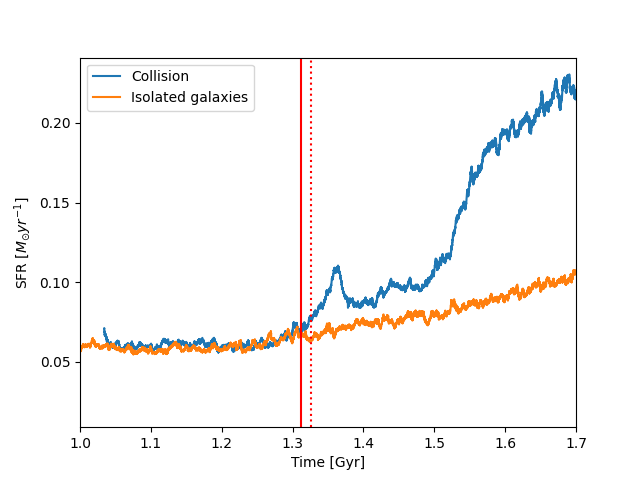}
    \caption{ Star formation rate evolution. Time is given in Gyr from the beginning of the simulation. The blue line shows the SFR in the encounter simulation. In orange we plot the sum of the SFR of the S0 and E galaxies when evolved isolated. Perigalacticum is at 1.311\,Gyr (vertical solid red line), where a sudden star formation episode begins. The dotted vertical line at 1.325\,Gyr marks the present state of the system, indicating that the SFR is rising due to the collision and will peak about 60\,Myr after perigalacticum.}
    \label{fig:sfr}
\end{figure}

 Fig.~\ref{fig:sfr} shows the time evolution around perigalacticum (at 1.311\,Gyr after the beginning of the simulation) of the star formation rate (SFR) during the simulation. Note that just after perigalacticum, a sudden peak in SFR occurs, as expected. We plotted in orange the SFR evolution of the isolated models to show that the peak is due to the collision.  At the moment, the system is undergoing an outbreak of star formation. In Section~\ref{sec:sps} we present a stellar population synthesis, from which we detect, in some apertures, and for both galaxies, the presence of a few percent of the young stellar population. The star formation rate peaks near 60\,Myr after perigalacticum, with an SFR of 0.12\,M$_\odot\,Myr^{-1}$.

Regarding the radial velocity distribution (Fig.~\ref{fig:velmap_sec}), it also corresponds nicely with the observed velocities over the slit positions. If we take the velocity profile along PA$=46.4^{\circ}$ from the velocity map, we will have a beautiful U-shaped profile, unlike what happens along the line that connects the nuclei of the galaxies.

\begin{figure}
\centering
	\includegraphics[angle=0,width=\columnwidth]{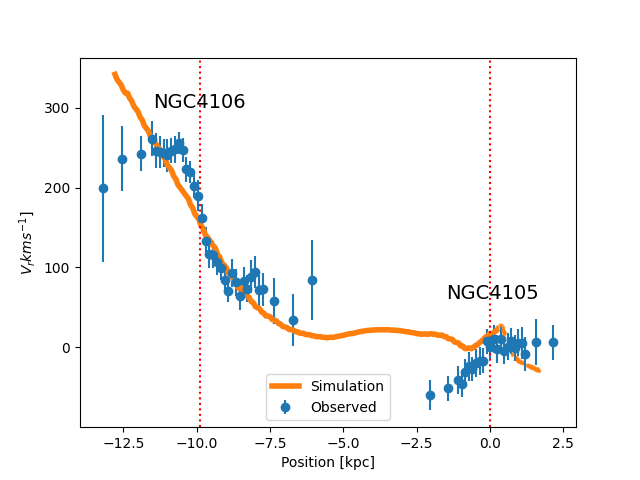}
   \caption{ Velocity profile along the PA\,=$118^{\circ}$. The observed data points are plotted in blue. In orange is a cut on the velocity map of the simulation (Fig. \ref{fig:velmap_sec}), over the same direction. The center of each galaxy is indicated by the dotted vertical red lines. }
    \label{fig:vel_prof_simul}
\end{figure}

The velocity profile along the PA=$118^{\circ}$ is presented in Fig. \ref{fig:vel_prof_simul}. The overall shape of the  velocity profile taken from the simulation is in good agreement with the observed one. At the left end of the graph, which corresponds to the SE side of the slit, the velocity in the simulation keeps rising to values above 300\,km\,s$^{-1}$, while the observed profile becomes flat at ${\sim}250$\,km\,s$^{-1}$. This is due to the details on that part of the simulated galaxy that develops a prominent bar, stronger than that on NGC\,4106. On the other side, the velocity profile in the NGC\,4105 section is quite strange, with a low velocity tail in the direction of its companion galaxy. It is clear that a simulation can hardly reproduce the observations in a very high degree of detail. Both simulations and observations have limiting factors that affect the data. The objective here is to study the dynamics in general lines.

\section{Stellar populations synthesis}
\label{sec:sps}

To compute the stellar population contribution on the galaxy pair, 
we utilized the 
{\scriptsize\,STARLIGHT} stellar population synthesis code \citep{2005MNRAS.358..363C}.
Full details on the code and methodology are discussed in 
\citet{2004MNRAS.355..273C,2005MNRAS.358..363C,2006MNRAS.370..721M,2007MNRAS.375L..16C,2007MNRAS.381..263A}.
Briefly, the code fits the observed spectrum (O$_{\lambda}$) in terms of the overlap of simple stellar population synthesis models (SSPs) 
with a wide range of ages and metallicities. In this code, we adopt the empirical stellar spectral library  by  
\citet{10.1046/j.1365-8711.2003.06897.x}. 
The library includes spectra with a spectral resolution of 3\,\AA{}\,(FWHM)  
over a wavelength range 
$\lambda\lambda3200\,-\,9500$\,\AA{} under a sampling of $1$\,\AA{}.
We also adopted the Padova 1994 tracks, and an initial mass function (IMF) by \citet{Chabrier_2003} for stars of masses 
between 0.1 and 100 M$_{\odot}$.

The basic equation of the synthetic spectrum solved by the {\scriptsize\,STARLIGHT} is given by:

\begin{equation}
\label{eq:synthetic}
\centering
M_\lambda/M_{\lambda_{0}} = \left[\sum_{j=1}^{N_\star} \vec{x}_j b_j,_{\lambda} r_{\lambda} \right] \otimes G(\nu_{\star},\sigma_{\star}),
\end{equation}

\noindent where \textit{M}$_{\lambda_{0}}$ is the synthetic flux at that rest-frame wavelength; {$\vec{x}_j$} is the population vector;
$b_{j}$,$_{\lambda}$ is the reddened spectrum of the $j^{\rm th}$ SSP model that is flux normalized at $\lambda_{0}=5750$\,{\AA};
$r_\lambda$\,$\equiv$\,10$^{-0.4(A_{\lambda}\,-\,A_{\lambda_{0}})}$ is the extinction term;
$\otimes$ represents the convolution operator;
G$(\nu_{\star},\sigma_{\star})$ denotes a Gaussian distribution along the line of sight centered at velocity $\nu_{\star}$
and with a star velocity dispersion $\sigma_{\star}$;
The population vector {\bf  $\vec{x}$} 
denote the
percentage contribution of the SSPs at $\lambda_{0}$ weighted by flux, being expressed in terms of age and metallicity ($t_{j}$, $Z_{j}$). 
The population vector can also be  expressed as a function of the SSP mass fractional contribution and is designed by 
the vector {\bf  $\vec{m}$}.

The goodness-of-fit between observed and synthetic spectra is derived from an 
algorithm that finds the results for the minimum value of ($\chi^{2}$).
The intrinsic reddening is modeled by the code as due to the light scattering by dust, 
adopting the extinction law of \citet{1989ApJ...345..245C}.
The SSPs considered in this work take into account
15 ages ($t = 0.001$, $0.003$, $0.005$, $0.01$, $0.025$, $0.04$, $0.1$, $0.3$, $0.6$, $0.9$, $1.4$, $2.5$, $5$, $11$ and $13$ Gyr), 
three metallicities ($Z$ = $0.2$, $1$, and $2.5 Z_{\odot}$), summing it up N$_{\star}=45$\,SSP components.

We  combined the individual components in  age bins following the prescription of 
\citet{2005MNRAS.358..363C} as young stellar population, $x_{y}$ or $m_{y}$ ($t\,<\,1\times10^{8}$ yr);
intermediate stellar population, $x_{i}$ or $m_{i}$ (1$\times10^{8}\leq\,t\,\leq1\times10^{9}$ yr), and old stellar population $x_{o}$ or $m_{o}$ ($t\,>1\times10^{9}$ yr). 
These components are employed to construct the SSPs with their flux ({$\vec{x}$}) and mass ({$\vec{m}$}) fractional contributions.

Another important point is that the {\scriptsize\,STARLIGHT} code usually
not provide the uncertainties of the resulting parameters. However, we estimated the uncertainties of the average parameters as 
a function of the spectrum quality indicated by signal to noise ratio (SNR) variations.  For this, we applied  20 levels of 
random flux perturbations over the central spectrum of each slit position observed, and  calculated the percentage differences 
between the new estimates and standard ones. These percentage differences were scaled to the other apertures  of the slit according 
to the SNR of the  spectrum considered. We considered only spectra with a good quality signal-to-noise of the continuum (at least 10).

\begin{figure*}
\centering
	\includegraphics[angle=270,width=\columnwidth]{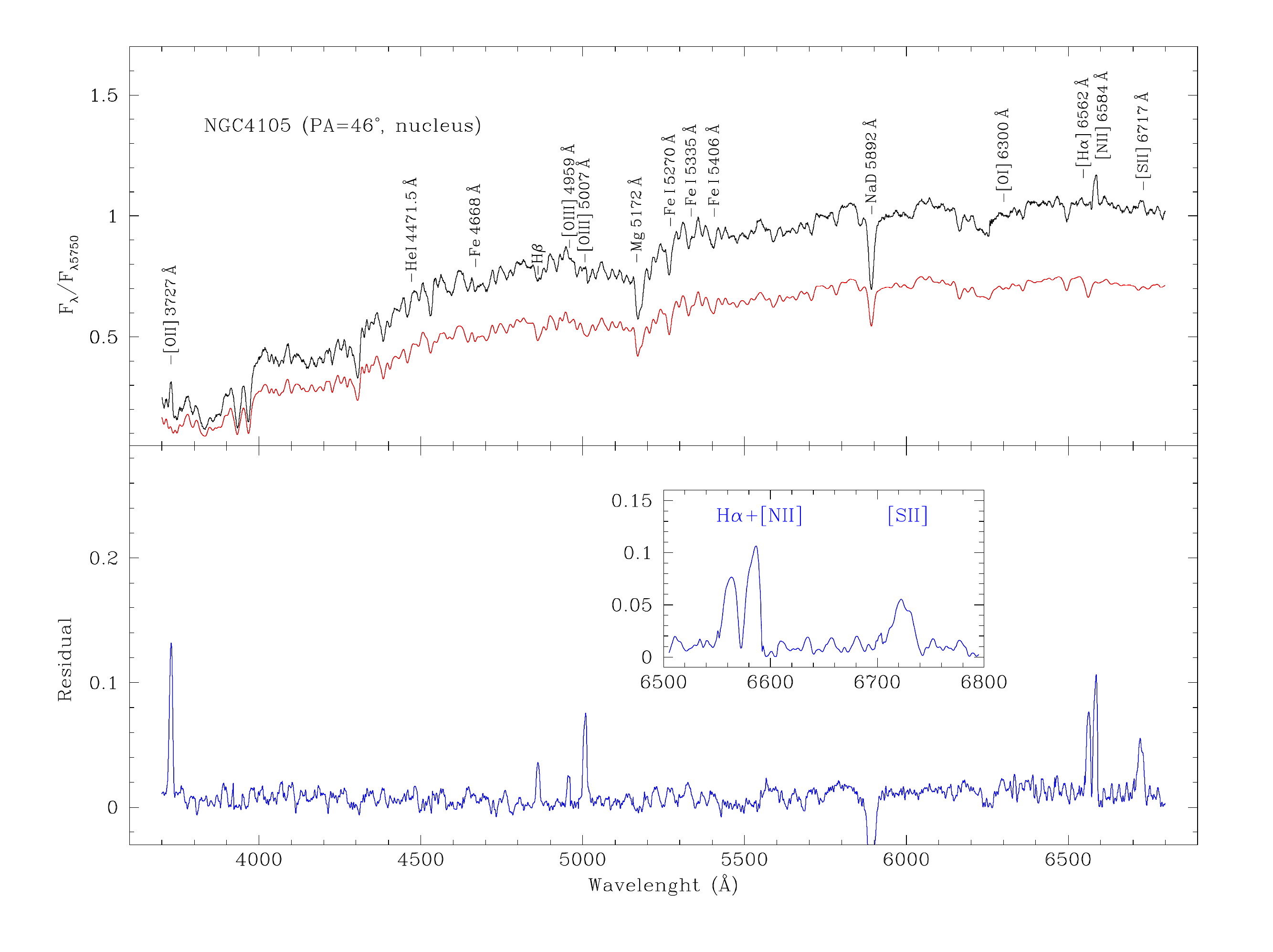}
	\includegraphics[angle=270,width=\columnwidth]{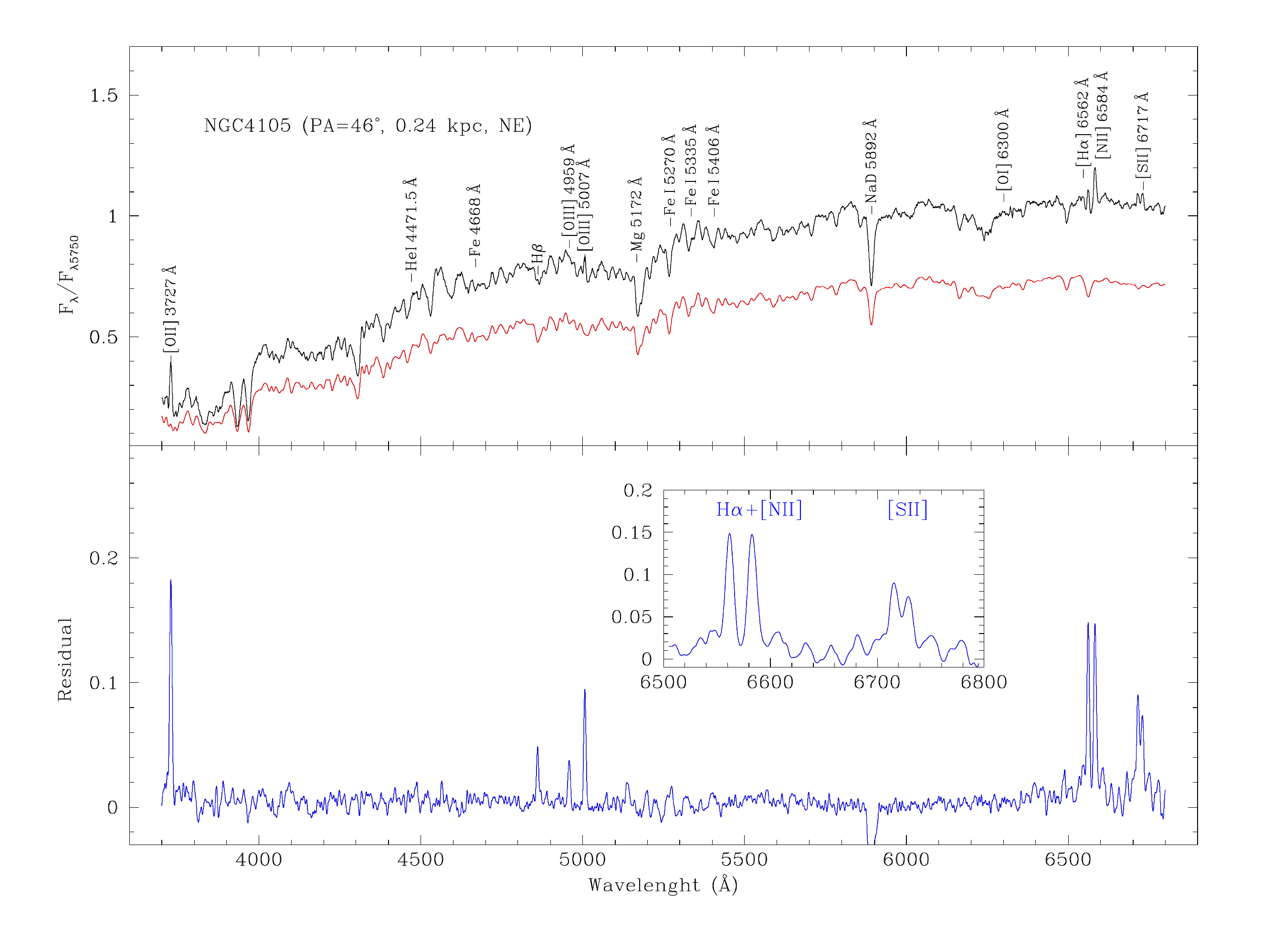}	
	\includegraphics[angle=270,width=\columnwidth]{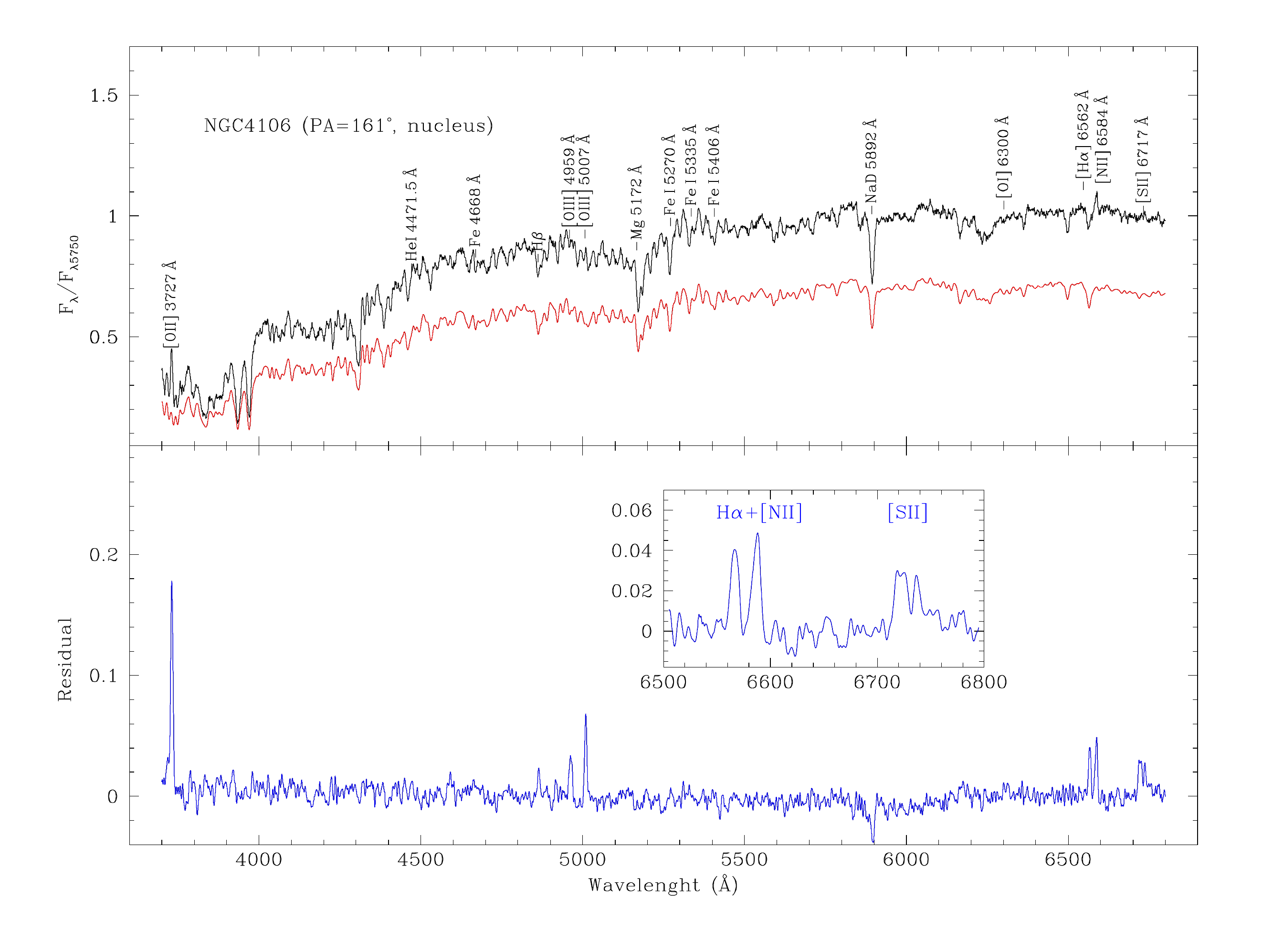}	
	\includegraphics[angle=270,width=\columnwidth]{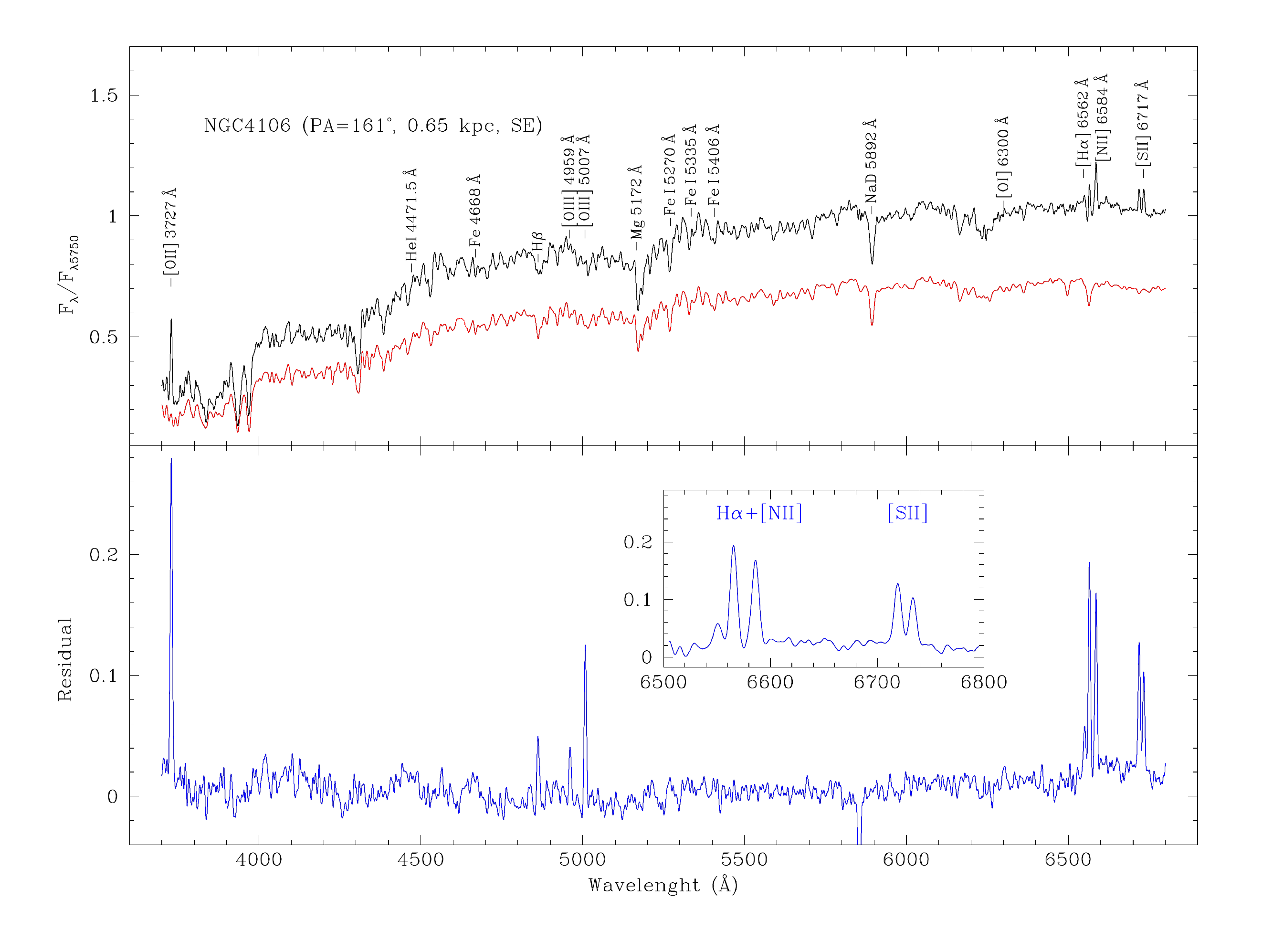}
    \caption{Stellar population synthesis for  different regions of NGC\,4105\,(top) and NGC\,4106\,(bottom).  The observed spectrum  (black),  the  synthesized spectrum (red), and the residual spectrum, with its pure emission line structure (blue), are shown. The main emission and absorption lines are identified, and a zoom-in of the region between 6500 to 6800 \AA{} is shown.}
    \label{fig:synth_espec}
\end{figure*}

\begin{figure*}
	\includegraphics[angle=0,width=\columnwidth]{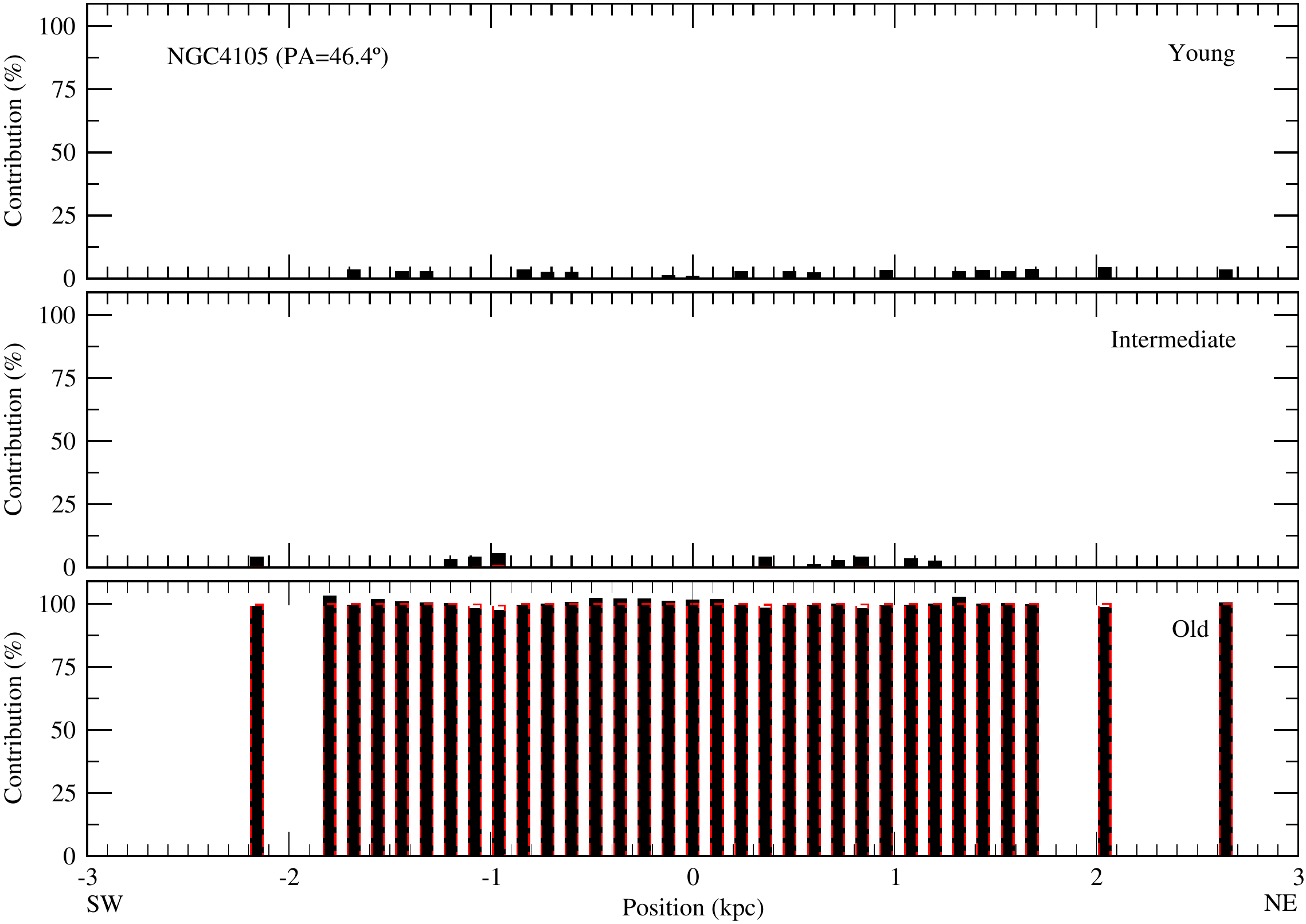}
	\includegraphics[angle=0,width=\columnwidth]{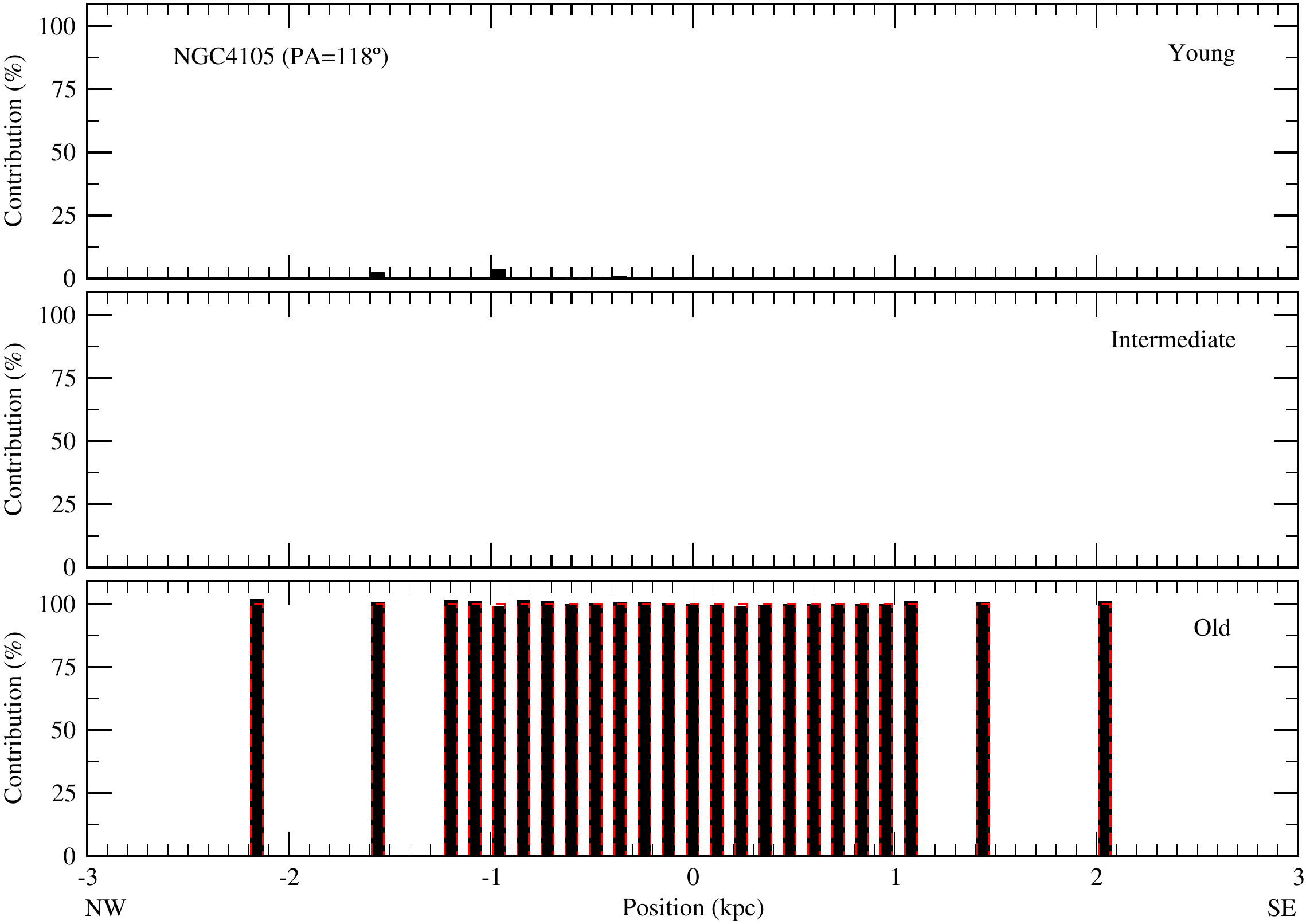}	
	\includegraphics[angle=0,width=\columnwidth]{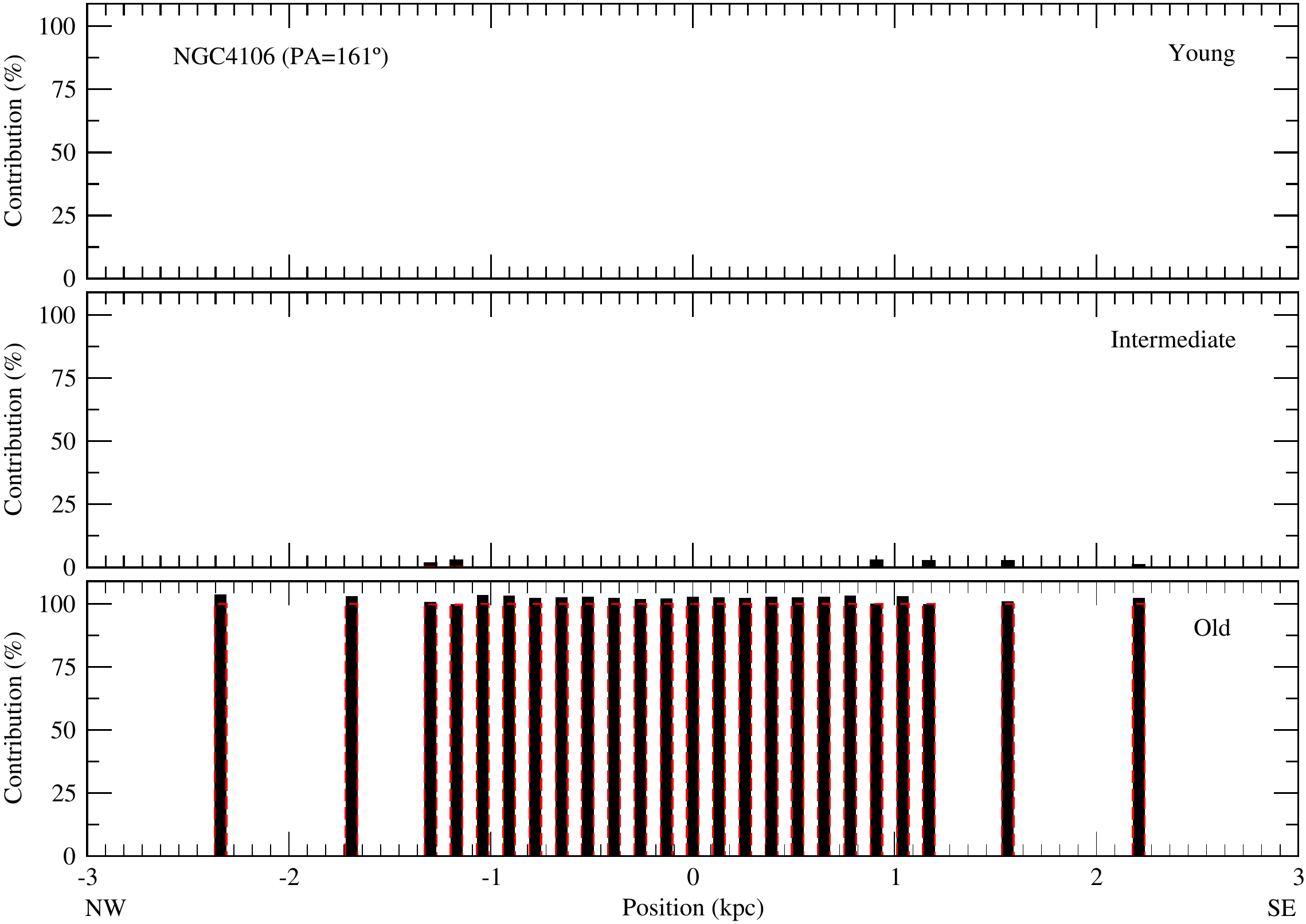}	
	\includegraphics[angle=0,width=\columnwidth]{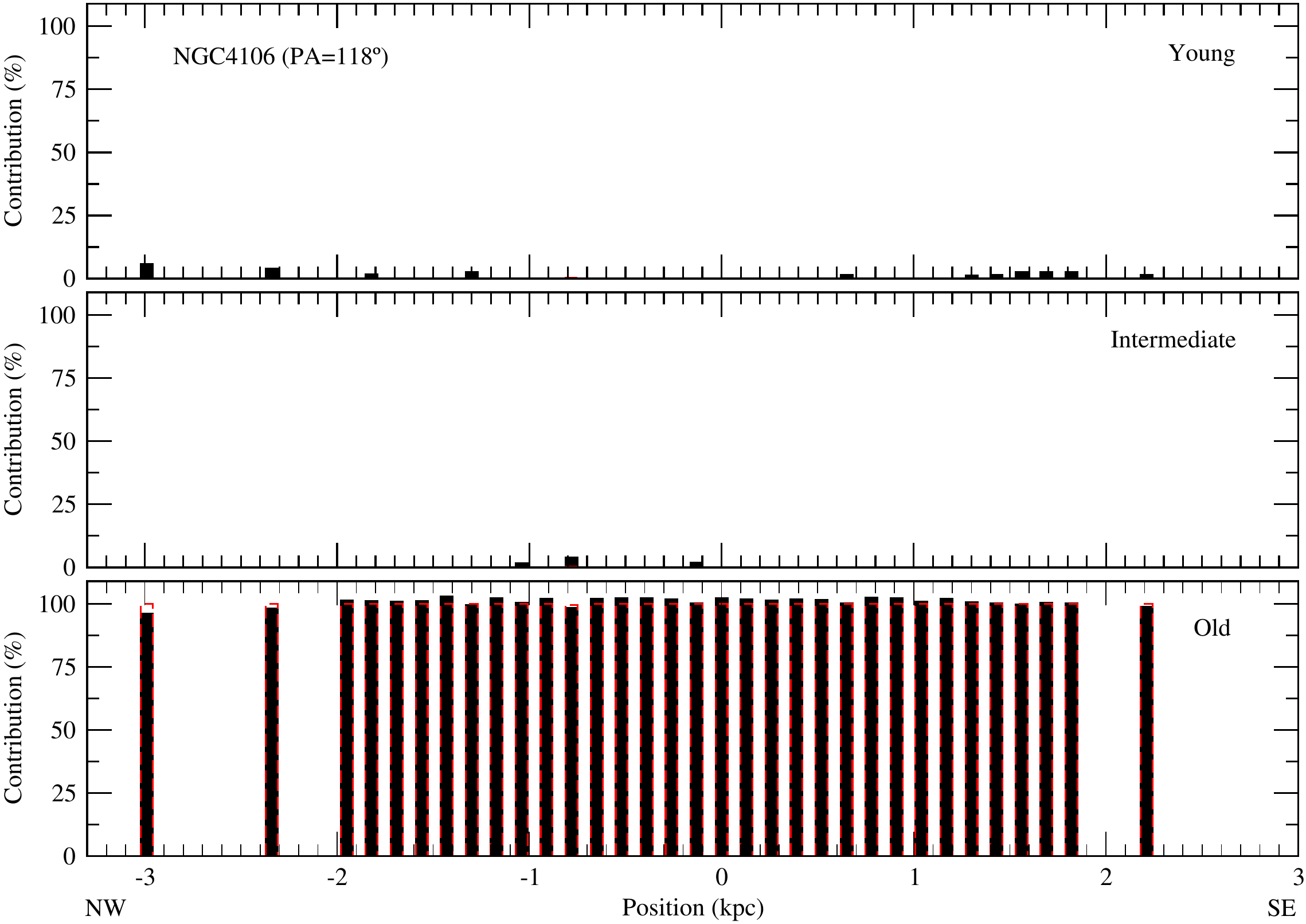}	
	\caption{Flux and mass fractions (black and red bars, respectively) of SSP contributions from the 
	best solution of {\scriptsize\,STARLIGHT} population synthesis.}
    \label{fig:synth_hist}
\end{figure*}

In Fig.~\ref{fig:synth_espec} we present some examples of the results  of the spectral synthesis fitting for the central and outskirt regions of NGC\,4105 and NGC\,4106. The results of stellar population synthesis for the individual spatial 
bins across the slit positions are shown in Fig. \ref{fig:synth_hist}, stated  as the fractional contribution of each base 
element weighted by flux and mass.
It can be noted that the spatial distributions of stellar populations are homogeneous across all slit positions. 
NGC\,4106 is a kind of lenticular galaxy,  so we would expect to find a light-weighted stellar age a few billion years younger than in the elliptical galaxy NGC\,4105. From the simulation, the maximum fraction of young stellar population ($\leq30$\,Myr) in NGC\,4105 would be 3.8\%. What we actually found for both galaxies was that the stellar population in both galaxies is predominantly composed by the old component in light, as well as in mass, and a very small fraction (up to 5 percent) of young and intermediate population, very close to the detection threshold of the stellar population synthesis method.

\begin{figure*}
\centering
\includegraphics[angle=0,width=0.9\textwidth]{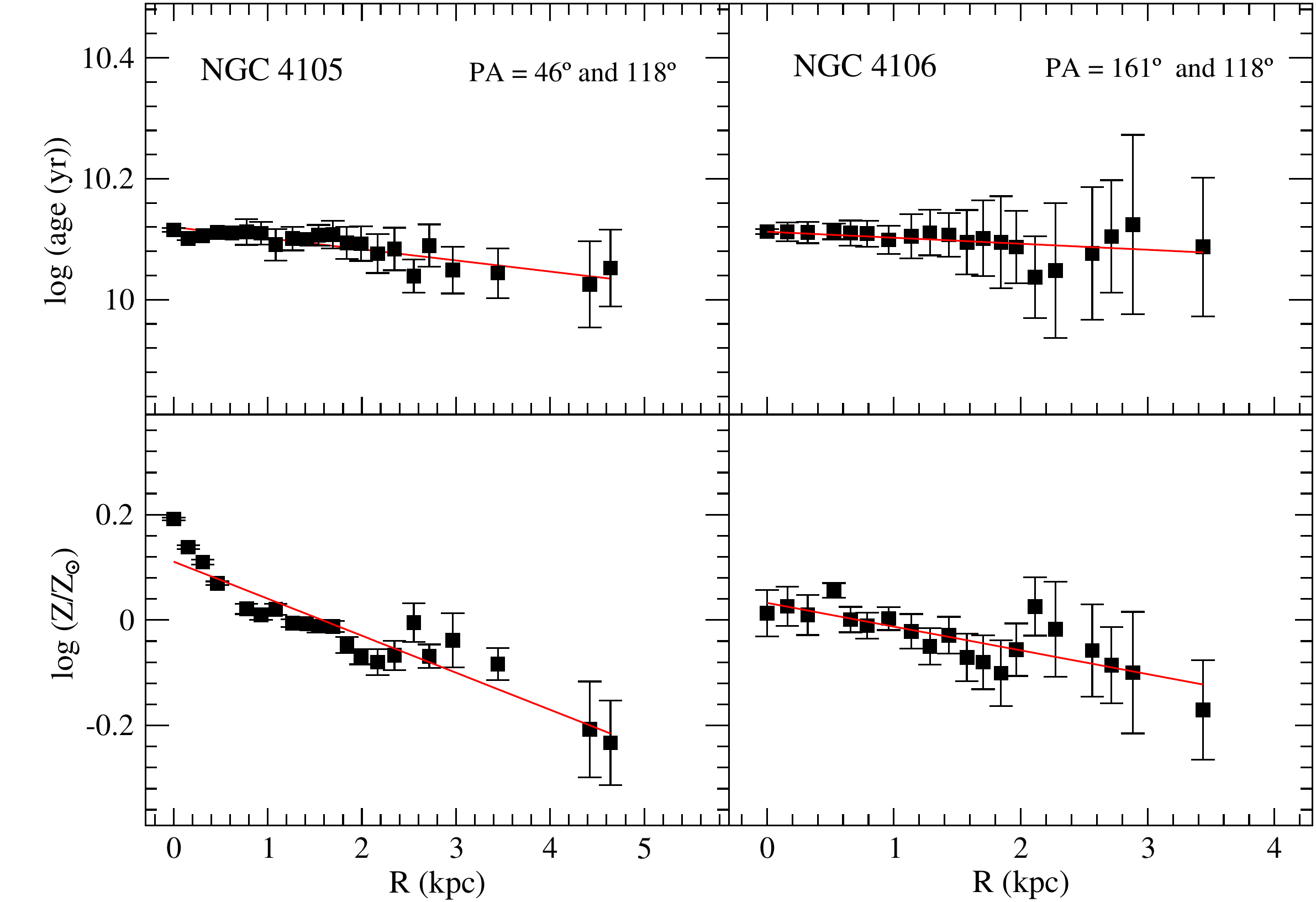}
\includegraphics[angle=0,width=0.9\textwidth]{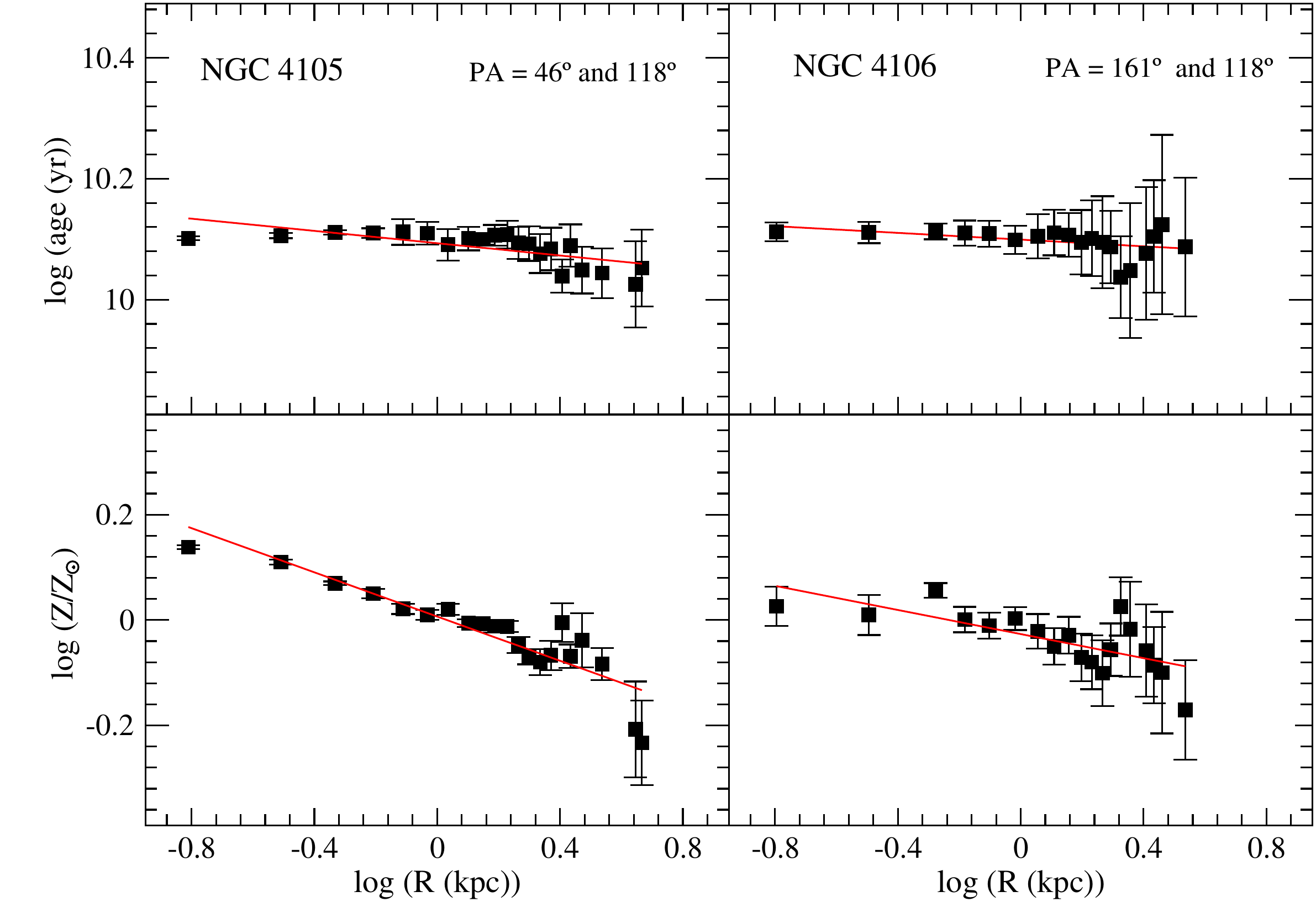}
\caption{The mean stellar age and metallicity (both in logarithmic scales)
as weighted by the flux contribution of SSP models addressed by the stellar population synthesis
as a function of the linear nuclear distance R (in kpc) and its logarithm, 
for NGC\,4105 and NGC\,4106.}

\label{fig:synth_grad}
\end{figure*}

We show, in Fig.~\ref{fig:synth_grad}, the light-weighted means of stellar age and metallicity
as a function of the linear distance towards the nucleus in each observed galaxy (expressed in kpc).
The age (in Gyr) is presented in the logarithm scale
and the metallicity Z in the logarithm scale normalized to the solar value, i.e. log\,($Z/Z_{\odot}$)\,=\,[Z],
for which $Z_{\odot}\,=\,0.02$ to be consistent with the adopted SSPs grid.
The stellar population synthesis has extracted light-weighted stellar age and metallicity
on two distinct projected directions of each galaxy,
across which we have also measured the stellar kinematics (see Subsection \ref{sec:cine_stellar}).
The stellar population properties have been derived from a set of aperture spectra
along radial distances up to almost 0.5\,R$_{\rm e}$ in the case of NGC\,4105,
and up to around 1\,R$_{\rm e}$ for NGC\,4106.
NGC\,4105 is morphologically classified as a E3 galaxy and NGC\,4106 as a SB(s)0+ galaxy,
i.e. a lenticular barred galaxy.
Specifically for NGC\,4106,
our kinematics observations have shown that the slit direction at PA\,=\,118$^{\rm o}$
is more aligned to a stellar rotating disc than one at PA\,=\,161$^{\rm o}$,
on which the maximum line-of-sight rotational velocity is
$128\pm22$\,km\,s$^{-1}$ and $50\pm21$\,km\,s$^{-1}$ respectively (see Fig. \ref{fig:cine_A}).
The other galaxy, the elliptical NGC\,4105, does not exhibit any stellar rotating disc.
Since the radial variations of both stellar properties
are indistinguishable across both directions in each observed galaxy,
we have aggregated them all for each object into a single dataset
in order to estimate representative radial gradients of light-weighted stellar age and metallicity.
The stellar properties derived for different apertures
at each side of the radial profile across two distinct projected directions
were averaged in the case they correspond to the same nuclear distance within 1\,arcsec,
otherwise, the apertures are individually considered.
The radial distances of all apertures towards the galaxy nucleus 
were corrected by the inclination angle
of the reference plane of each correspondent galaxy.

In the case of NGC\,4105 (a massive E3 galaxy),
we have measured, up to a nuclear distance of about 4.6\,kpc
(i.e., up to around 0.45\,R$_{\rm e}$),
a slightly negative gradient in age
($\Delta{\log}(\rm age\,(\rm Gyr))$/$\Delta{{\rm R(kpc)}}$\,=\,$-0.018\pm0.002$\,dex\,kpc$^{-1}$ and
$\Delta{\log}(\rm age\,(\rm Gyr))$/$\Delta{\log}({\rm R(kpc)})$\,=\,$-0.051\pm0.012$) 
and a very sharp negative metallicity gradient
($\Delta{\log}(\rm Z/Z_{\odot})$/$\Delta{{\rm R(kpc)}}$\,=\,$-0.070\pm0.006$\,dex\,kpc$^{-1}$
and $\Delta{\log}(\rm Z/Z_{\odot})$/$\Delta{\log}({\rm R(kpc)})$\,=\,$-0.210\pm0.023$).
Whilst the variation in stellar age is very small
(i.e., from nearly 13\,Gyr in the nucleus down to around 11\,Gyr at about 0.5\,R$_{\rm e}$),
the metallicity deeply decreases from about [Z]\,=\,+0.1\,dex in the nucleus
down to nearly [Z]\,=\,-0.2\,dex at about 0.5\,R$_{\rm e}$.

In the case of NGC\,4106 (an intermediate mass lenticular galaxy),
we have measured nearly up to 3.4\,kpc along the nuclear distance
(i.e., up to about 1\,R$_{\rm e}$)
a null gradient in age
($\Delta{\log}(\rm age\,(\rm Gyr))$/$\Delta{{\rm R(kpc)}}$\,=\,$-0.010\pm0.005$\,dex\,kpc$^{-1}$,
$\Delta{\log}(\rm age\,(\rm Gyr))$/$\Delta{\log}({\rm R(kpc)})$\,=\,$-0.028\pm0.015$) 
and a negative metallicity gradient
($\Delta{\log}(\rm Z/Z_{\odot})$/$\Delta{{\rm R(kpc)}}$\,=\,$-0.045\pm0.008$\,dex\,kpc$^{-1}$
and $\Delta{\log}(\rm Z/Z_{\odot})$/$\Delta{\log}({\rm R(kpc)})$\,=\,$-0.114\pm0.027$).
Whilst the stellar age is constant around 12.6\,Gyr inside 1\,R$_{\rm e}$ ($\pm$0.6\,Gyr),
the metallicity decreases from around the solar value in the nucleus (+0.03\,dex indeed)
down to nearly [Z]\,=\,-0.1\,dex at about 1\,R$_{\rm e}$ (-0.12\,dex indeed)
that closely corresponds to one third of the whole metallicity variation
on a two times greater normalized radial distance in comparison against NGC\,4105.

Negative gradients in stellar age and metallicity with different intensities
are observed in massive early-type galaxies (especially ellipticals)
\citep{2012A&A...538A...8S, 2015A&A...581A.103G, 2020MNRAS.491.3562Z},
such that older metal-richer stars are found within the innermost regions of the galaxy
and the relatively younger and metal-poorer stars in the outskirt ones,
as a result of a long two-phase process \citep{2010ApJ...725.2312O}.
The initial phase occurs at high redshift
and builds up the main body of the elliptical with stellar populations passively evolving afterwards.
In the second phase lasting billions of years, minor/major dry mergers make the main body progressively bigger and redder.
An extended relatively younger stellar envelope with relatively smaller metallicity than the nuclear region is also formed
\citep{2013ApJ...763...26O, 2016ApJ...821..114H}.
Very recently, \citet{2020MNRAS.491.3562Z} investigated and
compiled the spatial distribution of the light-weighted mean stellar age and metallicity
inside extended galactic regions (i.e., up to 2\,R$_{\rm e}$)
in a wide variety of nearby ETGs (48 E’s and 21 S0’s extracted
from the CALIFA integral field spectroscopic survey).
They found that the age profiles typically follow
a U-shaped variation in the plane log(age(Gyr))-R$_{\rm e}$
with a minimum around 0.4\,R$_{\rm e}$,
asymptotically increasing outwards beyond 1.5\,R$_{\rm e}$
and increasing towards the nucleus.
The greater the $\sigma_{\rm v}$ within 1\,R$_{\rm e}$ ($\sigma_{\rm e}$) is,
the smaller the depth of the minimum and the central increment are,
i.e., flatter the age gradient is.
The metallicity gradients are found universally negative and strong
such that the metallicity flattens out moving towards larger radial distances
(about -0.3\,dex per 1\,R$_{\rm e}$ within 1\,R$_{\rm e}$).
They analysed the metallicity profiles
over the plane ${\log}(\rm Z/Z_{\odot})$-${\log}({\rm R_{e}})$.
They state that a possible qualitative interpretation
for their observations is a two-phase scenario for the formation of ETGs.

In comparison with the age and metallicity radial gradients investigated by \citet{2020MNRAS.491.3562Z}
in the most massive elliptical galaxies (i.e., $\sigma_{\rm e}$\,$\geq$\,210\,km\,s$^{-1}$),
NGC\,4105 exhibits, within the central region up to 0.5\,R$_{\rm e}$,
a comparable change in age and a greater variation in metallicity.
In the case of NGC\,4106, this SB0 galaxy presents, within the central region up to 1\,R$_{\rm e}$,
smaller variations in both age and metallicity
in comparison against ETGs with intermediate mass (i.e., 170\,$\leq$\,$\sigma_{e}$\,$<$\,210\,km\,s$^{-1}$).
According to E’s against S0’s, \citet{2020MNRAS.491.3562Z} found
that they all have U-shaped radial profiles in age,
which are dislocated to smaller ages for S0’s by around 1.6\,Gyr, 
and comparable metallicity gradients within 0.5\,R$_{\rm e}$
that become distinct between 0.5 and 1\,R$_{\rm e}$ (steeper in E’s).

Since the nearest approximation between the galaxies seems to have happened only 30\,Myr ago,
we suppose the gradients of age and metallicity in both galaxies have not been yet substantially modified by the interaction in course.
Both galaxies might have preserved their intrinsic stellar population spatial distributions
according to their particular mass assembly histories.
In fact, \citet{2004MNRAS.347..740K} stated
that the merging histories of elliptical galaxies could, in principle, be extracted
from the observed metallicity gradients in the local Universe galaxies.
\citet{2004MNRAS.347..740K} obtained that there is no correlation between metallicity gradient and galaxy mass
based on a variety of chemodynamical simulations of elliptical galaxies (GRAPE-SPH code).
An average gradient in metallicity that \citet{2004MNRAS.347..740K} compiled was
$\Delta{\log}(\rm Z)$/$\Delta{\log}({\rm R/R_{e}})$\,=\,-0.3 ($\pm$0.2 as dispersion).
\citet{2004MNRAS.347..740K} also concluded that E galaxies that hypothetically would have been formed
through a monolithic process would have steeper gradients,
while E galaxies that undergo major mergers would have shallower gradients.

The stellar population synthesis has provided small flux contributions of young and intermediate age populations (respectively, with age $< 100$\,Myr and 100\,Myr\,$\leq$ age $\leq 1$\,Gyr)
in both galaxies, i.e., up to about 6 per cent, which is very close to the detection limit
of the stellar population synthesis applied through the {\scriptsize\,STARLIGHT} code.
Therefore, these very small contributions did not represent any significant perturbation on the age and metallicity radial gradients. However, \cite{2007AJ....133..220G} argue that NGC\,4106 shows a value for the H+K (Ca{\scriptsize\,II}) line-strength index larger than 1.3, i.e., larger than the maximum value attainable in post star-burst models (with both solar and supersolar metallicity), which suggests the presence of H$\epsilon$ in emission, considered a good indicator of recent star formation. The detection of [O{\scriptsize\,II}]$\lambda$3727, 3729 emission in the nucleus of NGC\,4105 is also suggestive of a recent star formation episode.

On the one hand, in strongly interacting galaxies,
in which there was some reservoir of cold gas to form new stars during a long period of time, 
the metallicity gradient becomes significantly flatter
than the ones observed in isolated galaxies, as shown for instance by 
\citet{2008MNRAS.389.1593K, 2010ApJ...721L..48K, 2011MNRAS.416...38K, 2009ApJ...695..580B, 2014MNRAS.444.2005R}.
The shallow metallicity gradients found in strongly interacting galaxies could be explained
as the interaction-induced gas flows from the outer parts to the centre of each component
\citep{1972ApJ...178..623T, 2007ApJ...658..941D}.
On the other hand, the observed galaxy pair under a strong interaction too,
taking into account that the galaxies likely had a single encounter in a few tens of million years ago
and there was no great available amount of cold gas,
none significant star formation was actually induced inside both components by the mutual interaction.

\begin{figure}
\centering
\includegraphics[angle=0,width=\columnwidth]{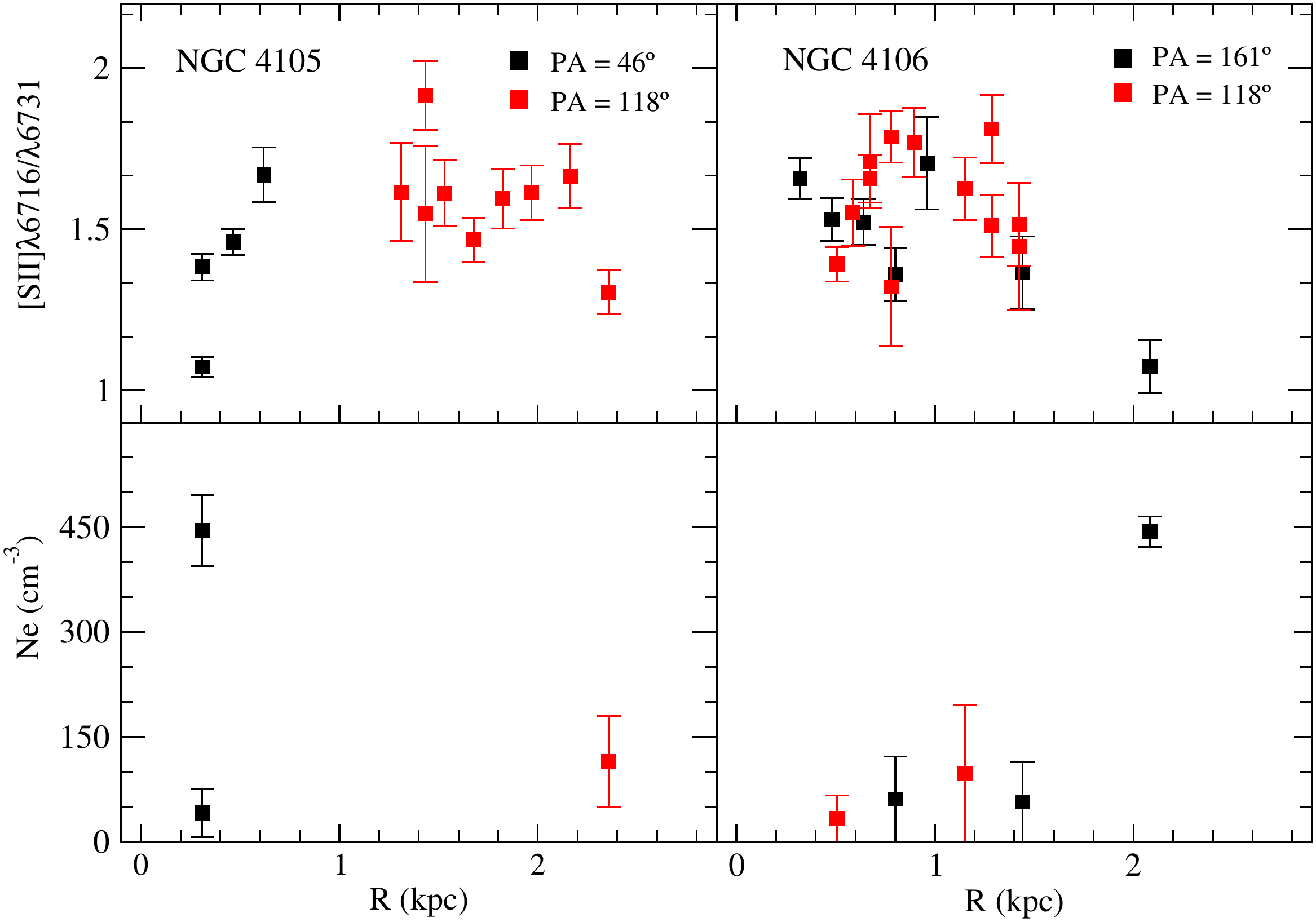}	\caption{
[S{\scriptsize\,II}]$\lambda\lambda$6716/6731\AA{} line ratio and electron density ($N_{\rm e} $)
as a function of linear radial distance in kpc.}
\label{fig:density}
\end{figure}

\section{Ionized gas properties}
\label{sec:gas_proprie}

The presence of emission lines are not unique  to star forming galaxies,  but may also appear in early-type galaxies 
\citep[e.g.,][]{1996A&AS..120..463M, 2000ApJS..127...39C, 2018Ap&SS.363..131R, 2018RNAAS...2....3D}.
Several studies have shown that line
emission is prevalent in more than 50\% of passive red galaxies
\citep{1986AJ.....91.1062P, 1994A&AS..105..341G, 2006ApJ...648..281Y, 2011A&A...529A.126C},
and many of the early-type emission
line galaxies show characteristic spectra of a LINER \citep{1980A&A....87..152H}. As can be seen in Fig.~\ref{fig:synth_espec}, after the subtraction of the modelled stellar spectrum from the observed one, the pure nebular spectra in both galaxies present pronounced emission lines,  that are:   [O{\scriptsize\,II}]$\lambda$3727, H$\beta$, [O{\scriptsize\,II}]$\lambda\lambda$4959,5007,
[N{\scriptsize\,II}]$\lambda\lambda$6548,6584, H$\alpha$, 
and [S{\scriptsize\,II}]$\lambda\lambda$6716,6731. In fact, previous studies have already detected some emission lines the elliptical galaxy NGC\,4105. For example, for NGC\,4105  \citet{1998A&AS..130..267L} detected  [O{\scriptsize\,II}]$\lambda$3727,  and \citet{2000ApJS..127...39C} measured [N{\scriptsize\,II}]$\lambda$6584.

The line intensities were determined from fitting Gaussian profiles on the pure emission spectra, 
using the {\scriptsize\,SPLOT} task in the {\scriptsize NOAO.ONEDSPEC} package of {\scriptsize\,IRAF}. 
The associated error with each  line flux was given as  {\scriptsize\,$\sigma^{2}=\sigma_{\rm cont}^{2}+\sigma_{\rm line}^{2}$}, 
where the  $\sigma_{\rm cont}$ and $\sigma_{\rm line}$ are the  continuum $rms$ and Poisson error on the line flux, respectively. We corrected the observed line intensities for the effect of interstellar extinction. 
This was performed comparing the observed  H$\alpha$/H$\beta$ ratio  with the  theoretical value of $2.86$ from \cite{1964MNRAS.127..145P}, 
for an electron temperature of 10\,000 K.
The  starburst extinction law of \citet{1994ApJ...429..582C} was adopted.

\subsection{Electron density}

We have determined the electron density  ($N_{\rm e}$) from  the  observed
[S{\scriptsize\,II}]$\lambda6716/\lambda6731$\AA\, ratio, 
using {\scriptsize\,TEMDEN} routine of the {\scriptsize\,NEBULAR} package of the {\scriptsize\,STSDAS/IRAF}, assuming an electron 
temperature of $10^{4}$\,K, since the lines sensitive to temperature were not observed in both galaxies. 
It is important to note that some regions in both galaxies present [S{\scriptsize\,II}] ratio near the 
low-density limit, and therefore, the electron densities for these areas cannot be estimated.

The radial profiles of [S{\scriptsize\,II}] line ratio and $N_{\rm e}$  
are plotted in Fig.~\ref{fig:density} as a function of the linear radial distance
for NGC\,4105 and NGC\,4106.  
The values of electron density obtained for both galaxies are in the range of $N_{\rm e}=33-445\,\rm cm^{-3}$ . 
These  estimations  
are in agreement with those obtained by  \citet{2014MNRAS.437.1155K}  and  \citet{2019MNRAS.488..830M}  for H{\scriptsize\,II} 
regions of interacting galaxies, which  are systematically higher than those derived for isolated galaxies of 
$N_{\rm e}=40-137\,\rm cm^{-3}$, as previously estimated  by \citet{2014MNRAS.437.1155K}.

\subsection{Oxygen Abundance}
\label{sec:bpt}

The identification of the dominant ionization source for the emitting gas across the galaxy is essential to determine the 
chemical abundance correctly. The elliptical galaxy, NGC\,4105, is classified as a LINER galaxy by \citet{2007MNRAS.375..931M} 
and \citet{2010A&A...518A..10V}, and NGC\,4106 does not have a classification in the literature. 
Therefore, with the goal of  confirming the ionization source of NGC\,4105 and to classify NGC\,4106, we used the  
$[\ion{O}{iii}]/\rm H\beta$ versus $[\ion{N}{ii}]/\rm H\alpha$ 
diagnostic diagram  proposed by \citet{1981PASP...93....5B}, and commonly known as BPT diagram. This diagram is used 
to distinguish objects  ionized by massive stars (SF), AGN, and LINER natures. In our case, we can not use the $[\ion{O}{iii}]/\rm H\beta$ versus $[\ion{S}{ii}]/\rm H\alpha$
because the   [S{\scriptsize\,II}]$\lambda\lambda$6716,6731 emission lines presented a very low S/N in the central regions of the galaxies.

\begin{figure}
\centering
    \includegraphics[angle=0,width=\columnwidth]{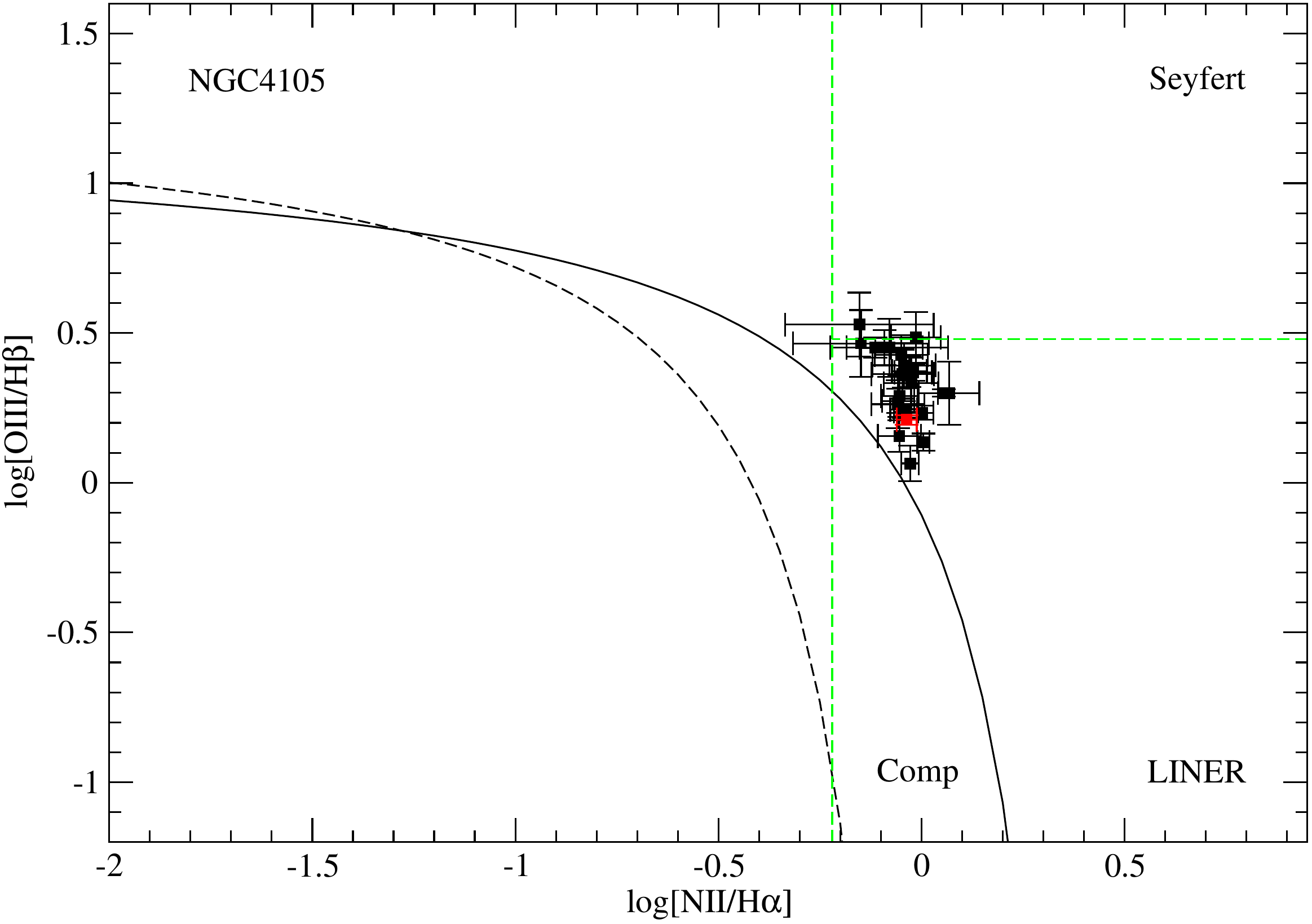} 
    \includegraphics[angle=0,width=\columnwidth]{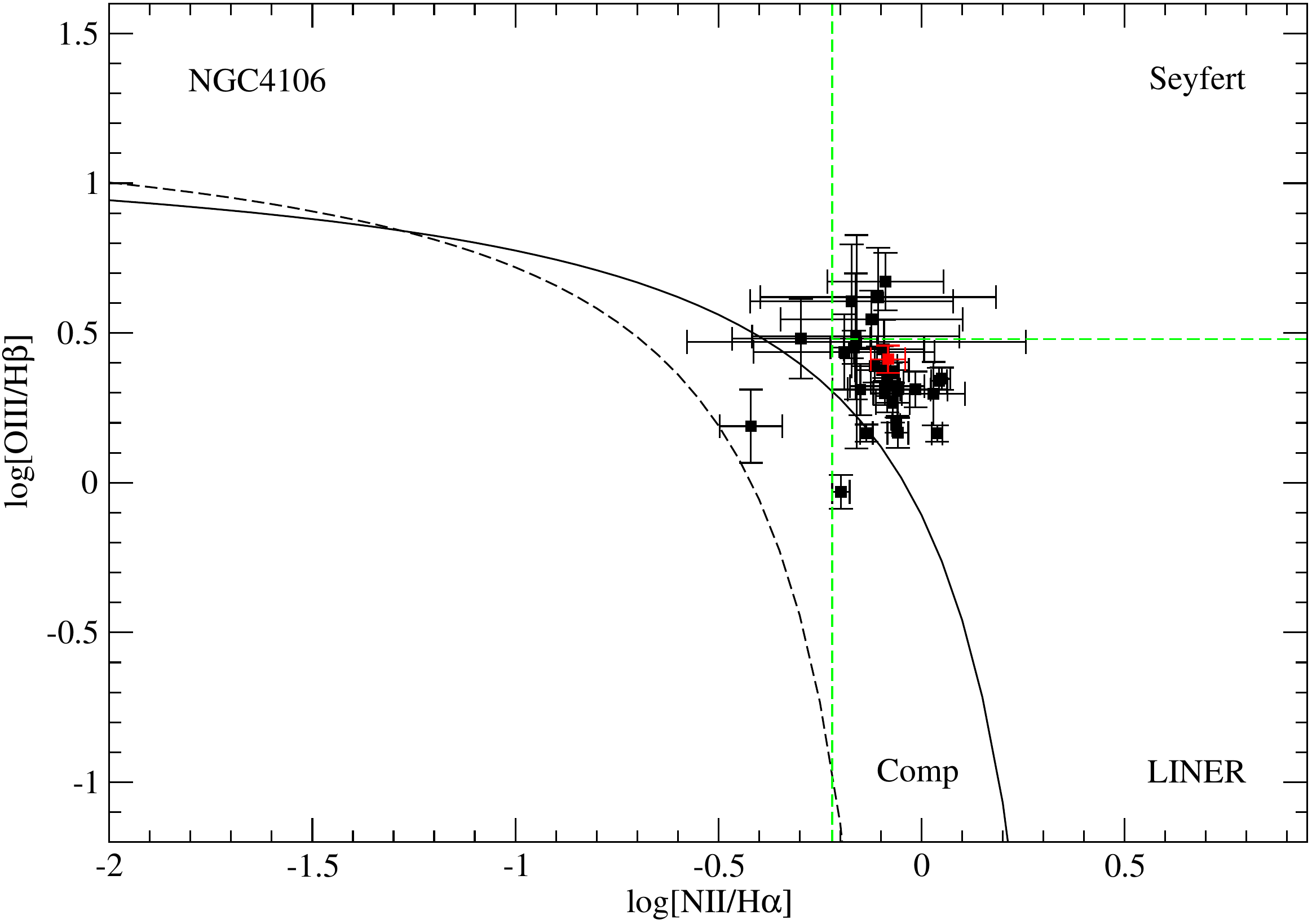} 
\caption{The  [O{\scriptsize\,III}]/H$\beta$  versus [N{\scriptsize\,II}]$\lambda$/H$\alpha$
diagnostic diagram. 
The central region of each galaxy is marked by red symbols.
The  solid  black  line   represents the  theoretical  upper  limit  for  SF  galaxies from \citet{2001ApJ...556..121K} (Ke01), 
and the black dashed curve pure star-formation line from \citet{2003MNRAS.346.1055K} (Ka03). 
The region between Ke01 and Ka03 is denominated composite region. 
 The dividing line between Seyferts and LINERs (long dashed, green) was set by  \citet{1997ApJS..112..315H}.}
\label{fig:bpt}
\end{figure}

Fig.~\ref{fig:bpt} present the BPT diagram for the regions measured  along the different slit positions of AM\,1204-292. 
We can note that all regions of NGC\,4105 present emission like LINER, in agreement with the findings of \citep{1991S&T....82Q.621D}.  NGC\,4106 shows extended LINER-type emission in their disc, as well as their  nuclei.

The accurate determination   of the metallicity is critically dependent on the emission lines sensitive to the electron temperature, 
which  are not detected in the spectra of the objects.  Therefore, to 
estimate the oxygen abundance of the gas phase in the nuclear regions of each galaxy, indirect methods that  use relatively 
strong emission lines easily observed in AGNs or SFs and/or photoionization models  should be applied. 
The most common method used for estimating the oxygen abundance  use the 
$R_{23}$=([O{\scriptsize\,II}]$\lambda$3727+[O{\scriptsize\,III}]$\lambda\lambda$4959,5007)/H$\beta$ parameter 
initially proposed \citet{1979MNRAS.189...95P}.
We used the $R_{23}$ indicator to estimate the metallicity 
in the nuclear regions of both galaxies, 
comparing the observed values with a 
grid of photoionization models from the code 
{\scriptsize\,CLOUDY} code \citep{2017RMxAA..53..385F}.
 
The grids were built following the same methodology presented in \citet{2014MNRAS.443.1291D,2017MNRAS.468L.113D}.
We created theoretical models based on power laws of the form $f_{\nu} \propto \nu^{\alpha}$, where $\alpha$ 
was assumed to be equal to $\alpha=-1.4$, which is  representative value for AGNs \citep{1981ApJ...245..357Z,2011ApJ...726...20M}, 
with  metallicities of $Z=0.2, 0.5, 1.0,2.0,3.0\,$and$ \,4.0\,Z_{\odot}$,
ionization parameter ${\rm log}\,U=-3.46,-2.46,1.46,0.46,0.54,1.54\,$and$\,2.54$,
and a fixed electron density value of 500\,cm$^{-3}$.
The solar oxygen abundance of $12+{\rm log(O/H)}_{\odot}=8.69\pm0.05$ is taken from \citet{2001ApJ...556L..63A}.

\begin{figure}
\centering
\includegraphics[angle=0,width=\columnwidth]{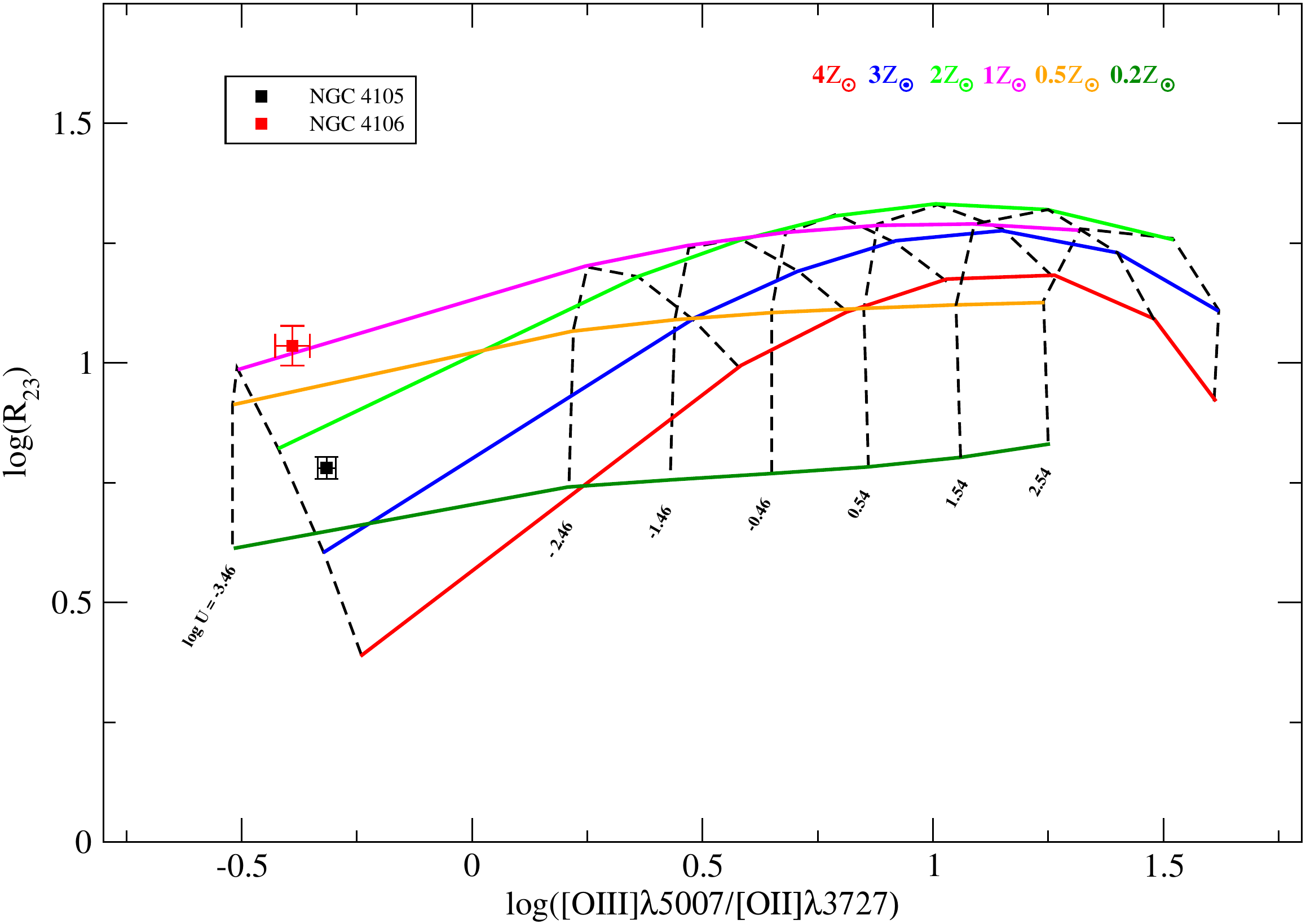}	
\caption{log(R$_{23}$) vs. log([O{\scriptsize\,III}]/[O{\scriptsize\,II}]) index.
The dashed  lines connect the photoionization model curves with different values of the logarithm ionization parameters, while the solid lines to different gas-phase metallicity. Black and red squares correspond to the nuclear region in NGC\,4105 and NGC\,4106, respectively.}
\label{fig:photoionization}
\end{figure}

Fig.~\ref{fig:photoionization} shows the $R_{23}$ vs. [O{\scriptsize\,III}]/[O{\scriptsize\,II}] diagram, with the observed values 
superposed in the computed  models.
Black and red squares correspond to the nuclear aperture in NGC\,4105 and NGC\,4106, respectively.
The central O/H abundances to each object were obtained by linear interpolation from the model grid, which 
resulted in  $12+{\rm log(O/H)}=9.03\pm0.02$ and  $12+{\rm log(O/H)}=8.69\pm0.05$ for NGC\,4105 and NGC\,4106, respectively. 
We compared the  metallicities obtained from the gas and stellar phase and verified that for NGC\,4106 the $Z/Z_{\odot}$ is the same, that is, 
 $Z/Z_{\odot}= 1$, and for NGC\,4105, the metallicity in the stellar phase ($Z/Z_{\odot}= 1.6$) is slightly lower than that obtained 
 from the gas $Z/Z_{\odot}= 2.2$, about 30\%.

\section{Conclusions}
\label{sec:conclusions}

With the present paper, we aimed at understanding the effects of the interaction in the 
stellar and gas kinematics, stellar populations and ionized gas properties in the strongly interacting system pair AM\,1209-292. Long-slit spectra in the range $3000-7050$\,{\AA}  were obtained with the Goodman High Throughput Spectrograph attached to the $4.1-$m SOAR telescope.
The main results are the following:

\begin{enumerate}

\item No signs of rotation were detected in the stellar components over both slit directions observed for NGC\,4105.  As for gas, the kinematics show a clear pattern of rotation for PA=$118^\circ$, that is, gas and stars do not share the same kinematics. We can deduce from this that the ionized gas in NGC\,4105 is of external origin and that it must have been incorporated into the galaxy recently. However, it was not the result of the ongoing interaction with NGC\,4106, as the modeling of the system dynamics indicates. In contrast, NGC\,4106 shows nearly symmetrical rotation curves across both slit directions, and a dynamical mass of $1.6\times10^{10}\,M_{\odot}$ was calculated from the V$_{0}$\,=\,138$\pm$6\,km\,s$^{-1}$ at a radius of 3.64 kpc.
  
\item  We also perform a set of N-body numerical simulations of the encounter between both components using P-Gadget3 TreePM/SPH code. It was found that the perigalacticum would have occurred  about 14.2\,Myr ago. At this point, the system is undergoing an outbreak of 
star formation. In addition, our models successfully reproduce some important observational features of the system. 

\item The contribution of the stellar components in relation to the optical flux at $\lambda5870$ \AA{} for both galaxies  is dominated by old ($t\,>1\times10^{9}$ yr) population, with a not negligible contribution of an intermediate population and a 
small amount of young population.
 
\item The electron density estimates for the pair AM\,1204-292 are systematically higher than those derived for isolated galaxies in the literature. Some regions of  NGC\,4106 and NGC\,4105 galaxies show an increment of $N_{\rm e}$ towards the outskirt  and  in  the close central region, respectively.
  
\item We observe a LINER-like line ratio in all regions of NGC\,4105. While NGC\,4106 shows extended LINER-type emission in the disc, as well as their nuclei.

\item The central O/H abundances to NGC\,4105 and NGC\,4106 were obtained by linear interpolation from a model grid and intensity of emission lines, which resulted in  $12+{\rm log(O/H)}=9.03\pm0.02$ and  $12+{\rm log(O/H)}=8.69\pm0.05$ for NGC\,4105 and NGC\,4106, respectively. 

\item We compared the  metallicities obtained from the gas and stellar phase and verified that for NGC\,4106 the $Z/Z_{\odot}$ is the same, that is, $Z/Z_{\odot}= 1$, and for NGC\,4105, the metallicity in the stellar phase ($Z/Z_{\odot}= 1.6$) is slightly lower than that obtained from 
 the gas $Z/Z_{\odot}= 2.2$, about 30\%.

\end{enumerate}

In summary, the current work has contributed to the understanding of how the mutual dynamic perturbation in a strongly interacting galaxy pair can disturb the global morphology of each galaxy, the stellar and gas kinematics, the physic-chemical properties of the gaseous component, 
and the stellar formation inside each one of the galaxies. The AGN phenomenon can also be intensified by encounters between galaxies containing gas in some amount, having been investigated in this work as well. Curiously, the system contains galaxies of different morphological types (and with distinct masses too), such that one object could have been wrongly classified as a spiral galaxy due to the formation of a double tidal stellar tail during the interaction as proposed by us, which has easily been misinterpreted as a spiral arm. This kind of observational study on a detailed characterization of both stellar and gaseous components, when confronted with a set of N-body numerical simulations for the encounter, has provided a lot of interesting results about the formation and evolution of interacting galaxies in very low density environments like nearby galaxy pairs and sparse groups.

\section*{Acknowledgements}

D. A. Rosa is grateful for the scholarship from CAPES and CNPq foundations for the PCI/MCTIC/INPE posdoc fellowship (process 300082/2016-9). I. Rodrigues and A. Krabbe thank the Brazilian foundation CNPq/MCTIC (grants number 313489/2018-1 and 311331/2017-3, respectively). A. Krabbe thanks FAPESP (grant number 2016/21532-9). Numerical simulations were run on the {\it Hypercubo} HPC cluster at IP\&D-UNIVAP (FINEP 01.10.0661-00, FAPESP 2011/13250-0, FAPESP 2013/17247-9 and FAPESP 2014/10489-0). We all acknowledge the usage of the HyperLeda database (http://leda.univ-lyon1.fr). This research made use of the NASA/IPAC Extragalactic Database (NED), which is operated by the Jet Propulsion Laboratory, California Institute of Technology, under contract with the National Aeronautics and Space Administration. {\scriptsize\,IRAF} is written and supported by the National Optical Astronomy Observatories (NOAO) in Tucson, Arizona. NOAO is operated by the Association of Universities for Research in Astronomy (AURA), Inc. under cooperative agreement with the National Science Foundation. We would like to thank Louis Ho, Zhao-Yu Li, and the CGS team for kindly providing the  image in the $I$ band of AM\,1204-292. 

\section*{Data Availability}
The data underlying this article will be shared on reasonable request to the corresponding author.



\bibliographystyle{mnras}
\bibliography{publicacao}


\bsp	
\label{lastpage}
\end{document}